\documentclass[aps, prd, print, 11pt]{revtex4-2}
\usepackage{hyperref}
\usepackage{bigints}
\usepackage{amsmath, amssymb, mathtools, sansmath, enumitem, braket}
\usepackage{graphicx}
\usepackage{verbatim}
\usepackage{dcolumn}
\usepackage{bm}
\usepackage{cancel}
\usepackage{mathtools}
\usepackage[normalem]{ulem}
\usepackage{tikz-feynman}
\usepackage[normalem]{ulem} 
\usepackage{xcolor} 

\newcommand{\ud}{\mathrm{d}}

\usepackage{subcaption}

\begin{document}
\preprint{APS/123-QED}
\title{Matter Sourced Bubble Nucleation in the Asymmetron Scalar-Tensor Theory}
\author{Usama Syed Aqeel}
\author{Clare Burrage}%
\author{Oliver Gould}
\author{Paul M. Saffin}
\affiliation{Centre for Astronomy \& Particle Theory, The School of Physics \& Astronomy, University of Nottingham, University Park, Nottingham, NG7 2RD, UK}%
\date{\today}

\setlength{\parindent}{0cm}
\begin{abstract}
We investigate how matter density distributions affect thin-wall bubble formation in the asymmetron mechanism, a scalar–tensor theory with a universal coupling to matter and explicit symmetry-breaking, and analyse the stability of its metastable state. We show that the screening mechanism of the asymmetron inside dense objects induces a surface tension associated with the boundary of the screening object, leading to a richer class of bubble solutions than the standard Coleman–Callan bulk nucleation. These boundary surface tensions are used to modify the Nambu-Goto action for instantons, allowing for the computation of the corresponding Euclidean action for bubbles nucleating on flat planes, as well as on concave and convex cylindrical surfaces. We find that the smallest Euclidean action occurs for bubbles nucleating along the edge of a concave spherical surface. Comparing this edge nucleation channel with the bulk one, we determine the maximum curvature radius for which concave edge nucleation is preferred. Since the maximum radius of curvature is exponentially suppressed by the action of a bulk bubble, we find that within the regime of the instanton approximation, edge nucleation is always preferred. This is largely due to the weak couplings of the asymmetron. We apply these findings to determine the maximum curvature radius of a cosmic void and discuss how our results affect the seeding of $N$-body simulations of asymmetron domains, showing that domain wall nucleation preferentially occurs at the edges of cosmological voids. We also demonstrate that the presence of a homogeneous gas around the dense substrates reduces the maximum curvature radius, enabling bulk bubbles to form preferentially as the asymmetron undergoes a density-driven phase transition. 
\end{abstract}

\maketitle
\newpage
\section{\label{sec:Introduction}Introduction}
The dynamics of scalar fields in the early Universe provide essential insights into cosmological phase transitions \cite{Mazumdar_2019}, vacuum stability \cite{PhysRevD.15.2929, PhysRevD.16.1762, PhysRevD.21.3305, PhysRevLett.46.388} and the origin of cosmic structure \cite{PhysRevD.23.347, LINDE1983177, 1992PhR...215..203M}. In particular, non-perturbative Euclidean solutions of the field equations, known as instantons \cite{Paranjape:2017fsy, Coleman:1985rnk}, describe tunnelling from a metastable vacuum to the true vacuum of a theory.

$ $

A well-studied application of scalar fields in Cosmology is the \textit{symmetron mechanism} introduced in Ref \cite{PhysRevLett.104.231301}, a scalar-tensor theory in which the scalar has a bare symmetry-breaking potential. In the Einstein frame, the scalar is minimally coupled to gravity, while in the Jordan frame, it couples universally to Standard Model matter fields. The presence of matter shifts the effective mass of the symmetron at the origin, making it tachyonic in low-density environments and real in regions of high density. Consequently, the effective potential has a single minimum in the high-density environment of the early Universe, but as the matter density redshifts below a critical density, the $\mathbb Z_2$ symmetry spontaneously breaks and the potential develops two minima \cite{Hinterbichler:2011ca}. As the scalar rolls into the different minima in distinct patches of the Universe, a domain wall network forms. A crucial feature of the symmetron in the cosmological picture is that it delays the formation of domain walls to a redshift of $z\sim 1$, as mentioned in Ref \cite{Hinterbichler:2011ca}, which prevents the energy density of the Universe from becoming dominated by the domain wall network during the inflationary epoch \cite{vilshell94, Vachaspati:2006zz}, which ensures the symmetron model is consistent with current cosmological observations in Ref \cite{DESIDR2II2025}.

$ $

The more general asymmetron potential \cite{peri, farbod, Chen_2015} introduces a cubic term that explicitly breaks the $\mathbb Z_2$ symmetry of the bare potential. The effective potential features false and true vacua \cite{PhysRevD.15.2929}. As in the symmetron, the onset of symmetry-breaking at low densities generates domains separated by unstable walls \cite{peri}. The pressure difference from the true domain onto the false domain destabilises the network. The earliest systematic treatment of metastable decay was given by J.~S.~Langer in Ref.~\cite{LANGER1969258, LANGER1967108}, who developed the theory of nucleation of first-order phase transitions in thermodynamic systems. Building on the world of Langer, Coleman and Callan \cite{PhysRevD.15.2929, PhysRevD.16.1762} established the semiclassical method for computing the associated decay rate, which remains a cornerstone of modern studies of vacuum stability. A pedagogical review of false vacuum decay and modern applications beyond bubble nucleation can be found in Ref.~\cite{Devoto_2022}.

$ $

Recently, there has been a rise in interest in seeding false vacuum decay in analogue condensed matter systems \cite{PhysRevD.110.105015, jenkins2025bubblesboxeliminatingedge}. Semiclassical analyses suggest that nucleation may be enhanced at boundaries, such as the edge of a cylindrical substrate \cite{PhysRevD.110.105015}. However, Ref.~\cite{brown2025mitigatingboundaryeffectsfinite,jenkins2025bubblesboxeliminatingedge} argues that for analogue experiments to faithfully mimic cosmological tunnelling, boundary effects, such as bubble nucleation on the interior edge of a spherical chamber, should be suppressed, since bubble nucleation in the early Universe is a bulk process.

$ $

Seeded nucleation has also been considered in curved spacetimes. For example, the nucleation of scalar bubbles around black holes in Schwarzschild-de-Sitter (SdS) spacetimes has been explored in Refs.~\cite{PhysRevD.32.1333, Gregory_2014, Gregory_2020, Shkerin:2021zbf}. In such settings, decay of the false vacuum acquires a thermal interpretation, since the Euclidean continuation renders time compact \cite{Brown_2007}, which induces a continuation of the Euclidean action into a statistical thermal free energy of the scalar field. Additionally, spacetimes with horizons, such as de Sitter space, have an associated temperature known as the \textit{Gibbons-Hawking} temperature \cite{PhysRevD.15.2738}. However, spacetimes with multiple horizons, such as SdS, exhibit different surface gravities, implying different temperatures; thus, a global thermal equilibrium is absent, except in special cases, like the Nariai limit (where the temperatures of the cosmic and black hole horizons are equal and constant~\cite{PhysRevD.57.2436}). While these studies neglect the quantum effects of the black hole, such as black hole evaporation, the approximations can be justified on timescales shorter than the horizon, where non-equilibrium effects are negligible. 

$ $

In this work, we investigate how quantum fluctuations render the asymmetron false vacuum metastable: the scalar field can tunnel through the barrier, nucleating a spherical bubble of true vacuum. The decay rate is computed semiclassically from the Euclidean continuation of the path integral \cite{Coleman:1985rnk, PhysRevD.16.1762, Coleman:1985rnk, PhysRevD.21.3305, Paranjape:2017fsy}. We extend the calculations of Coleman and Callan to discuss the role that boundaries play in bubble nucleation. Unlike the discussion in Ref.~\cite{brown2025mitigatingboundaryeffectsfinite, jenkins2025bubblesboxeliminatingedge}, boundary nucleation is physically relevant here, since the explicit coupling of the asymmetron to dense astrophysical objects such as stars and the edges of galactic voids acts as an effective boundary for the scalar field. This raises the question of whether bubble nucleation is enhanced on such geometries, and under what conditions does edge nucleation dominate over bulk nucleation? Moreover, we are motivated by laboratory searches for symmetron-like fifth forces, which remain sensitive to parameter ranges in which these effects could be tested \cite{Llinares:2018mzl, Clements:2023bva}.

$ $

The remainder of this paper is organised as follows. In Sec. \ref{sec:Instantons}, we use the Nambu-Goto instanton action to calculate bubble nucleation rates in the presence of a constant background density. We then expand this in Sec.~\ref{sec:heterogeneous} to bubbles nucleating on planar and cylindrical boundaries. We identify convex and concave edge nucleation channels and show that concave edges minimise the Euclidean action. In Sec.~\ref{sec:discussion} we compare bulk and edge decay rates, determine the maximum curvature radius for which edge nucleation dominates, and apply this both to laboratory vacuum chamber geometries and to cosmological settings such as galactic voids.

\section{\label{sec:level1}Background}
\subsection{\label{sec:universal-coupling}Universal Coupling of Scalars to Matter}
\subsubsection{The Symmetron Mechanism}
The symmetron mechanism, first introduced in Ref.~\cite{PhysRevLett.104.231301}, is a screening mechanism for the addition of scalar degrees of freedom in modified theories of gravity, known as scalar-tensor theories \cite{Joyce_2015, Clifton_2012}. The symmetron has the following action,
\begin{equation}\label{symmetron-action}
\begin{split}
    S[g^{\mu\nu}, \phi] &= \int \ud^4 x\sqrt{g}\left[\frac{1}{2}M_\text{p}^2R-\frac{1}{2}g^{\mu\nu}\nabla_\mu\phi\nabla_\nu\phi-V(\phi)\right]\\
     & \ \ \ \ \ \ \ \ \ \ \ \ \ \ \ \ \ \ \  +\int\ud^4x\sqrt{\widetilde{g}}\ \mathcal{L}_\text{m}(\psi_i, \widetilde{g}_{\mu\nu}) 
\end{split}
\end{equation}
where $g_{\mu\nu}$ and $\widetilde g_{\mu\nu}$ are respectively the Einstein and Jordan frame metrics, and $R$ is the Ricci scalar in the Einstein frame. $M_\text{p}$ is the Planck mass in the Einstein frame. The first line in Eq.~(\ref{symmetron-action}), expressed in the Einstein frame, is familiar, as it describes the theory of a scalar field that is minimally coupled to gravity. The second line is the action of the Standard Model, which is manifestly expressed in the Jordan frame. The two metrics are related to each other by the following Weyl transformation,
\begin{equation}
    \widetilde{g}_{\mu\nu} = A^2(\phi)g_{\mu\nu}.
\end{equation}
We can see that $\phi$ universally couples to all matter fields, $\psi_i$, through the term $\sqrt{\widetilde g}\ \mathcal{L}_\text{m}(\psi, \widetilde g_{\mu\nu})$.
The equation of motion for the scalar field is given by,
\begin{equation}
    \begin{split}
        \Box \phi = \frac{\ud V}{\ud\phi} - A^3(\phi)A_{,\phi} \widetilde{T}
    \end{split}
\end{equation}
where $\widetilde{T} =\widetilde{g}^{\mu\nu}\widetilde{T}_{\mu\nu}$ is the trace of the energy-momentum tensor of the matter in the Jordan frame, 
\begin{equation}
    \begin{split}
        \widetilde T_{\mu\nu} = -\frac{2}{\sqrt{\widetilde{g}}}\frac{\delta S_m}{\delta \widetilde{g}^{\mu\nu}}.
    \end{split}
\end{equation}

\begin{figure*}[t]
    \centering
    \includegraphics[width=1\linewidth]{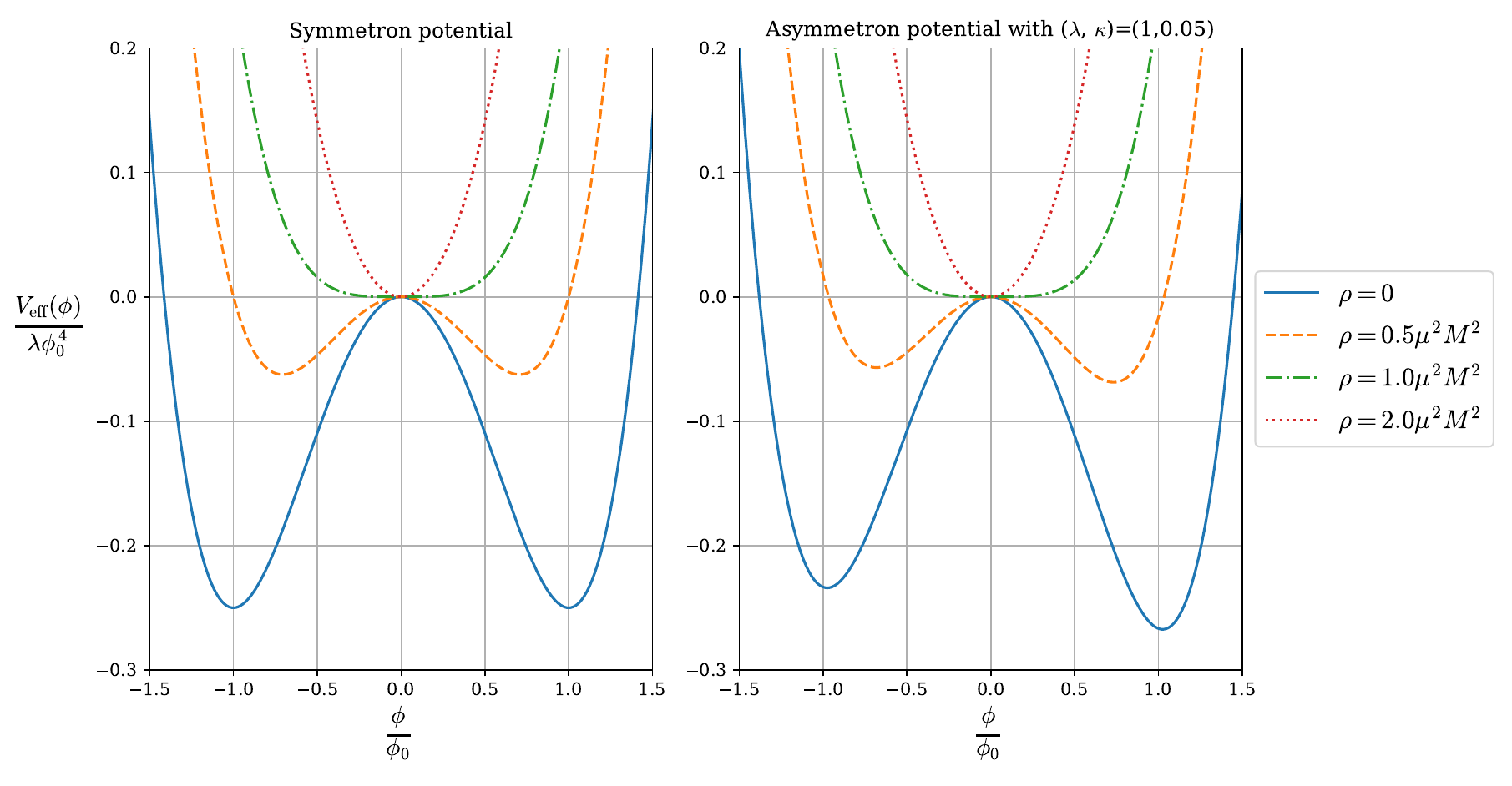}
    \caption{On the left, we show the symmetron effective potential from Eq.~(\ref{effective-potential}). On the right, we show the asymmetron effective potential Eq.~(\ref{effective-asymmeton-potential}) with parameter choices $(\lambda, \kappa) = (1,0.05)$. In both cases, it is observed that for densities $\rho < \mu^2M^2$, the potentials exhibit symmetry breaking and have two minima. We show the potential at the critical density $\rho = \mu^2M^2$, where it can be seen that the potentials acquire a single minimum. For densities larger than $\rho>\mu^2 M^2$, the potentials have a single global minimum. However, the asymmetron potential exhibits explicit symmetry breaking as the degeneracy-breaking in the potential in the right-hand figure is clear. For both potentials, we have rescaled the scalar field by the VEV of the bare potential, $\phi_0$, which allows us to factorise out some of the parameter dependence of the potential, leading to the prefactor, $\lambda\phi_0^4$, by which we scale $V_\text{eff}(\phi)$. In the symmetron case, this makes the potential only dependent on the local matter density $\rho$. In the case of the asymmetron, however, the ratio $\kappa/\lambda$ contributes to the size of the explicit symmetry-breaking of the minima.}
    \label{fig:potentials}
\end{figure*}

We will model the matter as a classical, pressure-free gas with density in the Jordan frame $\widetilde{\rho}$, in which case, $\widetilde{T} = -\widetilde{\rho}$. The corresponding Einstein frame density is given by, $\rho = A^3(\phi)\widetilde \rho$, and thus, the equation of motion for $\phi$ can then be written as, 
\begin{equation}
    \Box\phi = \frac{\ud}{\ud\phi}(V +\rho A) = \frac{\ud V_\text{eff}}{\ud \phi}
\end{equation}
where $V_\text{eff}(\phi)$ is the effective potential. The bare symmetron potential, $V(\phi)$, is chosen in Refs.~\cite{Hinterbichler:2011ca, Clements:2023bva, Llinares:2018mzl} to be the discrete $\mathbb{Z}_2$ symmetry-breaking potential given by, 
\begin{equation}\label{symmetry-breaking-potential}
    V(\phi) = -\frac{1}{2}\mu^2\phi^2 + \frac{\lambda}{4}\phi^4
\end{equation}
with minima given by $\phi_0 = \pm\frac{\mu}{\sqrt{\lambda}}$, where $\mu$ is the mass parameter and $\lambda$ is the quartic dimensionless coupling constant of the scalar. The symmetron is a theory for which we require the effective potential to have a $\mathbb{Z}_2$ symmetry and correspondingly, the Weyl transformation is required to respect the symmetry ($\phi\rightarrow-\phi$), 
\begin{equation}\label{Weyl-transformation}
    A(\phi) = 1+\frac{\phi^2}{2M^2} + \mathcal{O}\left(\frac{\phi^4}{M^4}\right)
\end{equation}
where $M$ is a cut-off scale that is introduced to ensure that $A(\phi)$ is dimensionless ($[\phi] = [M]$). This choice also reflects the fact that when $\phi$ vanishes, one retrieves Einstein gravity. It should be noted that the expansion in Eq.~(\ref{Weyl-transformation}) is valid for $\phi\ll M$ as discussed in Ref.~\cite{Burrage_2021}. Thus, the symmetron effective potential is given by,
\begin{equation}\label{effective-potential}
    \begin{split}
        V_\text{eff}(\phi) &= V(\phi) +\rho A(\phi) \\
        &= -\frac{1}{2}\mu^2\left(1-\frac{\rho}{\mu^2M^2}\right)\phi^2 + \frac{\lambda}{4}\phi^4
    \end{split}
\end{equation}
which is shown on the left-hand panel of FIG.~\ref{fig:potentials}. The symmetron mechanism is an example of a \textit{screening mechanism}, which is defined as a coupling of the scalar field to its environment which prevents the occurrence of long-range fifth forces. In the symmetron mechanism, the \textit{environment} is defined as the density distribution in spacetime. This explicit coupling to matter gives the symmetron an effective mass parameter $\mu_\text{eff}$ defined by, 
\begin{equation}\label{effective-mass}
    \mu_\text{eff}^2 = -\mu^2\left(1-\frac{\rho}{\mu^2M^2}\right).
\end{equation}
The scalar mass in Eq.~(\ref{effective-mass}) is \textit{tachyonic} ($\mu_\text{eff}^2<0$) when $\rho<\mu^2M^2$ resulting in1 spontaneous symmetry-breaking effects but, when $\rho>\mu^2M^2$, the mass parameter becomes real ($\mu_\text{eff}^2>0$) and the $\mathbb{Z}_2$ symmetry is restored. The spontaneous breakdown of symmetry in the symmetron model is driven purely by fluctuations in the local density of matter $\rho$. The \textit{critical density} of the symmetron potential is $\rho_*=\mu^2M^2$. We further define a region of \textit{subcritical matter density} when $\rho<\rho_*$ and a \textit{supercritical} one when $\rho>\rho_*$.

$ $

The symmetron potential in Eq.~(\ref{effective-potential}) admits domain wall solutions separating regions in distinct vacua that resemble the $\mathbb{Z}_2$ domain walls described in Ref.~\cite{Vachaspati:2006zz}. This is derived from the energy functional, $E[\phi]$, defined as, 
\begin{equation}
    E[\phi] = \int\left[\frac{1}{2}\dot\phi^2 + \frac{1}{2}(\nabla\phi)^2 + V_\text{eff}(\phi) \right]dV.
\end{equation}
We can use Bogomolnyi's method in Ref.~\cite{Bogomolny:1975de, Vachaspati:2006zz} to find the explicit profile and the associated energy. In three spatial dimensions $(x,y,z)$, the infinite, static planar domain wall $\phi=\phi(z)$ parallel to the $xy$-plane and centered at $z=0$ is given by,
\begin{equation}
    \begin{split}
        \phi(z) = \phi_0\left(1-\frac{\rho}{\rho_*}\right)^{\frac{1}{2}}\tanh\left[\frac{ z}{2 L_0}\left(1-\frac{\rho}{\rho_*}\right)^{\frac{1}{2}}\right].
    \end{split}
\end{equation}
When the density is supercritical, the domain walls vanish because the $\mathbb Z_2$ symmetry of $V_\text{eff}(\phi)$ is fully restored, resulting in a trivial vacuum manifold of the potential. We define $L_0=\left(\sqrt{2}\mu\right)^{-1}$ as the Compton wavelength of the symmetron. The domain wall, $\phi(z)$, carries a surface energy, $\sigma(\rho)$, which is localised around the core of the domain wall, given by,
\begin{equation}\label{energy-density}
\begin{split}
    \sigma(\rho)&=\int_{-\infty}^{\infty} \ud z \left[\frac{1}{2}\phi'(z)^2 + V_\text{eff}(\phi)\right] \\
    &= \int_{-\phi_0}^{\phi_0}d\phi\sqrt{2V_\text{eff}(\phi)}=\sigma\left(1-\frac{\rho}{\rho_*}\right)^{\frac{3}{2}}
\end{split}
\end{equation}
where $\sigma = \sigma(0)$ is the surface tension of the domain walls solution of the bare potential in Eq.~(\ref{symmetry-breaking-potential}),
\begin{equation}\label{domain-wall-vacuum-tension} 
    \sigma=\frac{2\sqrt{2}\mu^3}{3\lambda}
\end{equation}
discussed in Ref.~\cite{Vachaspati:2006zz}. On scales much larger than $L_0$, the domain wall appears to be a thin, flexible membrane, in which case, $\sigma(\rho)$ represents the surface tension, which is formally defined as the differential energy input, $dE$, required to increase the surface area of the domain wall by a differential amount $dA = dxdy$. 

$ $

For spatially-varying scalar field profiles, $A(\phi)$ contributes to a conservative potential for matter particles, leading to a fifth force, 
\begin{equation}
    \mathbf{F}_5  =-\mathbf{\nabla}(\ln A).
\end{equation}
This is most clearly demonstrated by taking the geodesic equation of a particle in the Jordan frame and expanding $\widetilde{g}_{\mu\nu}$ using Eq.~(\ref{Weyl-transformation}), which generates an additional tensorial term that survives when one moves into a local inertial frame. The dynamics of ultra-cold atoms interacting with symmetron domain walls in vacuum chamber experiments are discussed in Ref.~\cite{Llinares:2018mzl}, and there is a great interest in developing experiments, as in Ref.~\cite{Clements:2023bva}, to test for the existence of fifth forces. In high-density environments such as the solar system, the symmetron mechanism plays a crucial role in hiding these fifth forces \cite{Brax_2021, Panda_2024, Fischer_2024}. Local solar system tests constrain the parameter space ($\mu$, $\lambda$, $M$) of the symmetron such that the critical density is lower than the density of the solar system \cite{Clements:2023bva}. This means that the effects of these fifth forces should be below the sensitivity of current experimental tests of the Strong Equivalence Principle (SEP). 
\subsubsection{\label{sec:asymmetron}The Asymmetron Potential: Explicitly Breaking the Symmetron Degeneracy}
To construct an effective potential with explicit symmetry-breaking, one can either modify the bare potential $V(\phi)$ in Eq.~(\ref{symmetry-breaking-potential}) and/or the Weyl factor $A(\phi)$ in Eq.~(\ref{Weyl-transformation}). These modifications define a landscape of models with varied phenomenology. However, we are interested in a symmetron-like effective potential that exhibits explicit symmetry breaking of the minima in low-density environments. Consider the addition of a cubic term in $A(\phi)$ such as,
\begin{equation}
    A(\phi)=1+\frac{\phi^2}{2M^2} + \frac{\phi^3}{K}+\dots
\end{equation} 
breaking the $\mathbb Z_2$ symmetry in the Weyl factor. This generates $\widetilde{T}$-weighted corrections in the effective potential (e.g $\rho\phi^3/K$) and spoils the screening mechanism in high-density regions. Essentially, one would measure a non-zero scalar field value in regions of high density; therefore, the scalar asymmetron would register a fifth force that cannot be hidden from local solar system experiments. Higher-order odd terms would exacerbate the problem, resulting in an unbounded potential at large field values. Thus, we prefer to introduce a cubic term to the bare potential $V(\phi)$. This can be done most directly by adding a linear or cubic term to $V(\phi)$ without further modifications. Within the bare potential, it is always possible to remove a linear term by a shift $\phi\to\phi+\text{constant}$, at the expense of modifying the coefficients of the quadratic and cubic terms. Considering our asymmetron to be a low-energy effective theory, we avoid quintic or higher-order terms, which are non-renormalizable or \textit{irrelevant} in that their coefficients run to zero at low energies. We note that theories that derive from rescaling the metric, such as the Weyl transformation, will necessarily contain non-renormalizable operators. In the context of the Higgs metastability, which is induced by renormalisation group running, such non-renormalisable operators have been shown to have a sizeable influence on the decay rate and even on the existence of metastability~\cite{Eichhorn:2015kea, Branchina:2015nda, sodenheimer}. However, if metastability is present already in the tree-level potential, the effects of such non-renormalizable operators are expected to be smaller. We follow Refs.~\cite{peri, farbod} in adopting a cubic deformation of the bare potential $V(\phi)$,
\begin{equation}\label{bare-explicit-breaking-potential}
    V(\phi) = -\frac{1}{2}\mu^2
    \phi^2 + \frac{\lambda}{4}\phi^4 -\frac{\phi_0\kappa}{3}\phi^3
\end{equation}
where $\kappa$ is a dimensionless parameter such that $\kappa\ll\lambda$, $\phi_0$ is the VEV of the potential in Eq.~(\ref{symmetry-breaking-potential}) and is introduced in the coupling to make $\kappa$ dimensionless. Thus, we build the following effective potential for the asymmetron,
\begin{equation}\label{effective-asymmeton-potential}
    \begin{split}
        V_\text{eff}(\phi) = -\frac{1}{2}\mu^2\left(1-\frac{\rho}{\mu^2M^2}\right)\phi^2 + \frac{\lambda}{4}\phi^4 -\frac{\phi_0\kappa}{3}\phi^3
    \end{split}
\end{equation} 
shown in the right-hand panel of FIG.~\ref{fig:potentials}. Several properties of the asymmetron potential can be calculated exactly. We define a convenient factor $\Delta$, 
\begin{equation}
    \begin{split}
         \Delta^2(\kappa, \rho) =  1-\frac{\rho}{\mu^2M^2} + \left(\frac{\kappa}{2\lambda}\right)^2,
    \end{split}
\end{equation}
which appears frequently. The potential minima can be expressed as, 
\begin{equation}\label{exact-minima}
    \phi_\pm =\phi_0\left[\frac{\kappa}{2\lambda}\pm\Delta(\kappa, \rho)\right],
\end{equation}
where $\phi_+$ is the true vacuum (global minimum) of $V_\text{eff}(\phi)$ and $\phi_-$ is the false vacuum (local minimum), both of which are classically stable equilibrium states. The critical density discussed earlier also receives a correction from this cubic term, 
\begin{equation}\label{asymmetron-critdensity-correction}
    \rho_* = \mu^2M^2\left(1+\frac{\kappa^2}{4\lambda^2}\right),
\end{equation}
Since $\mathcal O(\kappa^2)$ terms are sub-leading, we will ignore them in the subsequent analysis; however, we shall acknowledge the existence of such corrections when they appear in important expressions.

$ $

The energy difference between the vacua is given by, 
\begin{equation}\label{energy-difference}
    \Delta V = V_\text{eff}(\phi_-)-V_\text{eff}(\phi_+) = \varepsilon\Delta^3.
\end{equation}
We denote the value of $\Delta V$ for $\rho=0$ and $\kappa\ll\lambda$ by $\varepsilon=\frac{2\mu^4}{3\lambda^2}\kappa$ and the leading-order contribution is given by $\Delta V = \varepsilon +\mathcal{O}(\kappa^3)$. The masses of small fluctuations around either vacua are given by, 
\begin{equation}\label{exact-masses}
\begin{split}
    m_\pm^2 = \left.\frac{\partial^2 V_\text{eff}}{\partial \phi^2}\right|_{\phi=\phi_\pm} = 2\mu^2\left[\Delta\pm \frac{\kappa^2}{4\lambda^2}\right]\Delta.
\end{split}
\end{equation}
If $\kappa \ll \lambda$, the mass splitting $\Delta m \equiv m_+-m_-$ is kept small and the masses about both vacua are approximately equal to
\begin{equation}\label{approximate-mass}
\begin{split}
         m &\equiv \frac{1}{2}(m_+ + m_-) = \sqrt{2}\mu\sqrt{1-\frac{\rho}{\mu^2 M^2}}\left[1 + \mathcal{O}\left(\frac{\kappa}{\lambda}\right)\right],
          \\
         \Delta m &\equiv m_+-m_- = \mathcal{O}\left(m\times\frac{\kappa}{\lambda}\right).
 \end{split}
 \end{equation}
Furthermore, we define the Compton wavelengths around each minimum, $L_\pm = m_\pm^{-1}$.

\subsection{\label{sec:Review-Instantons}Vacuum Decay} 
The early work of Langer in Refs.~\cite{LANGER1969258, LANGER1967108} established a framework for investigating the behaviour of thermal systems exhibiting metastable states. This was built upon by Coleman and Callan in Refs.~\cite{PhysRevD.15.2929, PhysRevD.16.1762}, to describe the behaviour of a zero-temperature quantum field theory system exhibiting explicitly-broken symmetry as discussed in Section \ref{sec:asymmetron}. In essence, they provide a generalisation of the Wentzel–Kramers–Brillouin (WKB) approximation to quantum field theory, yielding non-trivial semiclassical amplitudes associated with tunnelling processes. A detailed account of the mathematical framework of instanton calculations can be found in Refs.~\cite{Coleman:1985rnk, Paranjape:2017fsy}.

$ $

In the WKB methods for transitions between vacua of a potential $\phi_\text{i}\rightarrow\phi_\text{f}$ in a scalar field theory, one encounters the Euclideanized path integral given by,
\begin{equation}\label{path-integral}
    \begin{split}
        \bra{\phi_\text{f}}e^{-\frac{\beta H}{\hbar}}\ket{\phi_\text{i}} =\mathcal N\int \mathcal D\phi\exp\left(-\frac{S_E[\phi]}{\hbar}\right)
    \end{split}
\end{equation}
where $\mathcal N$ is an appropriately chosen normalisation constant. An instanton is a classical solution to the Euclidean equations of motion, derived from the Euclidean action, $S_E$, which itself is obtained by analytically continuing real time, $t$, to imaginary time $\tau$, via a Wick rotation,
$t\rightarrow-i\tau$ (see Ref.~\cite{Coleman:1985rnk, Paranjape:2017fsy}), 
\begin{equation}
    S\rightarrow -i S_E = -i\int \ud\tau \ud^3x\left[\frac{1}{2}(\partial_{\mu}\phi)(\partial_\mu\phi) + V(\phi)\right]
\end{equation}
Here, $(\partial_{\mu}\phi)(\partial_\mu\phi) = \left(\frac{\partial\phi}{\partial \tau}\right)^2 + (\nabla\phi)^2$ such that repeated indices downstairs are understood to mean a sum under a Euclidean signature and $\nabla$ refers to the spatial components of the gradient operator. The Euler-Lagrange equation is given by,
\begin{equation}\label{laplacian-instanton-equation}
    \partial_\mu\partial_\mu\phi = \frac{\partial^2 \phi}{\partial \tau^2} +\nabla^2\phi =  \frac{\ud V}{\ud\phi}.
\end{equation}
where $V(\phi)$ is a potential with an explicitly broken symmetry as with Eq.~(\ref{bare-explicit-breaking-potential}). The solution to this equation of motion, with appropriate boundary conditions, is the instanton. The path integral in Eq.~(\ref{path-integral}) can be evaluated using the saddle-point approximation, in which we expand the Euclidean action to second order around the instanton (the saddle-point of the Euclidean action),
\begin{equation}
    \left. S_E[\phi]\right|_{\phi_b} \approx S_E[\phi_b] + \frac{1}{2}\int d^4x \ \ell_n\phi_n^2 + \dots
\end{equation}
where $\phi_n$ are the normalised eigenfunctions of the second order fluctuation operator $S_E''[\phi]=-\phi[\partial_\mu\partial_\mu+V_\text{eff}''(\phi_b)]\phi$ around the bounce and $\ell_n$ are their corresponding eigenvalues.

$ $

It is shown in Ref.~\cite{Coleman:1977th} that in $D$ dimensions, a scalar field obeying the equation of motion in Eq.~(\ref{laplacian-instanton-equation}), the solution is $O(D)$ symmetric. In Refs.~\cite{PhysRevD.15.2929, Coleman:1985rnk} this spherically symmetric solution is called the \textit{bounce}, $\phi_b(\chi)$, where $\chi^2 = \tau^{2} + \sum_{i=1}^{D-2}x_i^2$ is the radial displacement from the origin. The Euclidean action of the $O(4)$-symmetric solution of Eq.~(\ref{scalar-bounce-equation}) can be expressed as,  
\begin{equation}\label{Spherically-symmetric-Euclidean-action}
    \begin{split}
        S_E[\phi] = 2\pi^2\int_0^\infty \ud\chi \ \chi^3\left[\frac{1}{2}\left(\frac{\ud\phi}{\ud\chi}\right)^2+ V(\phi)\right]. 
    \end{split}
\end{equation}
In four Euclidean dimensions, the equation of motion is given by,  
\begin{equation}\label{scalar-bounce-equation}
    \frac{\ud^2\phi}{\ud\chi^2} + \frac{3}{\chi}\frac{\ud\phi}{\ud\chi} = \frac{\ud V}{\ud\phi}.
\end{equation}
This resembles Newton's second law for a scalar field rolling down the inverted potential $V$ with a drag term inversely proportional to $\chi$. One often resorts to using the thin-wall approximation to solve equations of this form \cite{udemba2025quantumcorrectionssymmetronfifthforce, PhysRevD.15.2929}. The influence of the drag term is to force the scalar to remain near the true vacuum over a scale comparable to a radial distance $R\ll L_0$. Then, $\phi$ quickly rolls down the potential valley over a width that is much smaller than $R$ and quickly comes to rest on the false vacuum. One can use numerical methods to solve such equations away from the thin-wall approximation \cite{masters, Wainwright_2012}. Ref.~\cite{Matteini:2023ebf} discusses a perturbation series method to determine the bounce solution and its Euclidean action away from the thin-wall approximation using a quantity analogous to $\kappa$ as an expansion parameter. 

$ $

For a small energy difference between minima, $\kappa\ll \lambda$, the bounce is approximately given by, 
\begin{equation}\label{bounce-profile}
    \phi(\chi) =\begin{dcases}
        \phi_+ & \text{ for }(\chi\ll R)\\
        -\frac{\mu}{\sqrt{\lambda}}\tanh\left(\frac{\mu}{\sqrt{2}}(\chi-R)\right) & \text{ for }(\chi\sim R)\\
        \phi_- & \text{ for }(\chi\gg R)
    \end{dcases} 
\end{equation}
On this scale, the bounce has Euclidean action contributions that are proportional to the area and volume of a spherically symmetric profile. Like a bubble or a droplet separating the vapour and liquid phases, a scalar field bubble separates the true vacuum from the false vacuum.

$ $

The thin-wall bubble, with surface area $A$ and volume $V$, can be equally thought of as a closed hypersurface embedded in flat Euclidean spacetime $\mathbb R^4$ with a tension $\sigma$, defined in Eq.~(\ref{domain-wall-vacuum-tension}) on the bubble membrane and an outward-pointing pressure, $\varepsilon$, acting on the membrane, 
\begin{equation}\label{thin-wall-approximation}
    S_E = \sigma A -\varepsilon V.
\end{equation}
The question of finding the solution of lowest Euclidean action can now be answered using a geometric interpretation, where performing variations of Eq.~(\ref{thin-wall-approximation}) gives the equation of a constant mean curvature hypersurface $K=\frac{\varepsilon}{\sigma}$, where $K$ is the mean curvature of the bubble profile. In the thin-wall approximation, the bubble is a compact hypersurface of constant mean curvature - the only example of which is a sphere, $S^3$, for $D=4$.

$ $

The decay rate, $\Gamma$, of the false vacuum, derived in Ref.~\cite{PhysRevD.16.1762} is given by,
\begin{equation}\label{decay_rate}
    \begin{split}
        \Gamma = \mathcal V\times \left(\frac{S_E}{2\pi\hbar}\right)^2\frac{1}{\sqrt{|\ell_{-}}|}\left(\frac{\overline{\det}\ \mathcal{O}_b}{\det \ \mathcal O_-}\right)^{-\frac{1}{2}}\exp\left({-\frac{S_E[\phi_b]}{\hbar}}\right).
    \end{split}
\end{equation}
Here, $\mathcal O_b = -\partial_\mu\partial_\mu + V''(\phi_b)$ denotes the second-order fluctuation operator evaluated on the bounce solution $\phi_b$, while $\mathcal O_- = -\partial_\mu\partial_\mu + V''(\phi_-)$ is the corresponding operator around the false vacuum solution $\phi_-$. The symbol $\overline \det$ refers to the determinant of the positive-definite part of the spectrum, with the negative eigenvalue $\ell_-$ and the zero modes separated explicitly. The existence of a negative mode in the fluctuation spectrum is the signifier of the metastability of the false vacuum. The prefactor $\mathcal V \times \left(\frac{S_E}{2\pi\hbar}\right)^2$ arises from integrating over the zero modes of the spectrum using the method of collective coordinates, discussed in Ref.~\cite{PhysRevD.110.116023}, where $\mathcal V$ represents the infinite spatial volume of the Euclidean background. The Euclidean action, $S_E$, defined in Eq.~(\ref{decay_rate}) also appears in the exponential suppression factor - a result of a WKB approximation discussed in Refs.~\cite{PhysRevD.15.2929, Coleman:1985rnk}. A comprehensive derivation of these factors is presented in Ref.~\cite{Paranjape:2017fsy}.

$ $

An in-depth review of the thin-wall approximation applied to domain walls and instanton methods is presented in the series by M\'egevand \& Membiela in Ref.~\cite{Megevand:2023nin} with higher-order corrections included in subsequent papers \cite{Megevand_2024, Megevand_2025}.

\subsection{\label{sec:classical-nucleation-theory}Classical Bubble Nucleation}
We now turn to the nucleation of classical bubbles in the presence of \textit{substrates}. To address this, we adapt techniques from classical nucleation theory, focusing on bubble formation on surfaces (commonly referred to as \textit{heterogeneous nucleation}). We aim to investigate how density inhomogeneities affect the rate of bubble formation.

$ $

For a thin membrane separating two phases (true and false vacua separated by our bubble wall) to be stable, two conditions from hydrostatics must be satisfied simultaneously, found in Ref.~\cite{Soleimani}. These are: 
\begin{itemize}
    \item \textit{The Laplace condition}: the pressure difference ($\varepsilon$) and the surface tension ($\sigma$) across the membrane with $D-1$ principal radii of curvature obey the following relation, 
\begin{equation}\label{laplace-equation}
    \begin{split}
        \sum_{i=1}^{D-1}\frac{1}{R_i} = K
    = \frac{\varepsilon}{\sigma}
    \end{split}
\end{equation}
where $R_i$ are the principal radii of curvature and $K$ is the mean curvature of the membrane \cite{elementary-diff}. The bubble and substrate share a boundary $\Sigma\cap\mathcal{S} = \partial \Sigma$ which is a codimension-2 hypersurface that we shall call the \textit{contact hypersurface}. In FIG \ref{fig:planar-substrate}, the contact hypersurface is represented by a yellow loop (in $D=3$ this is referred to as the contact line).

$ $

\begin{figure*}[t]
\begin{subfigure}[t]{.5\textwidth}
  \centering
  \includegraphics[width=0.6\linewidth]{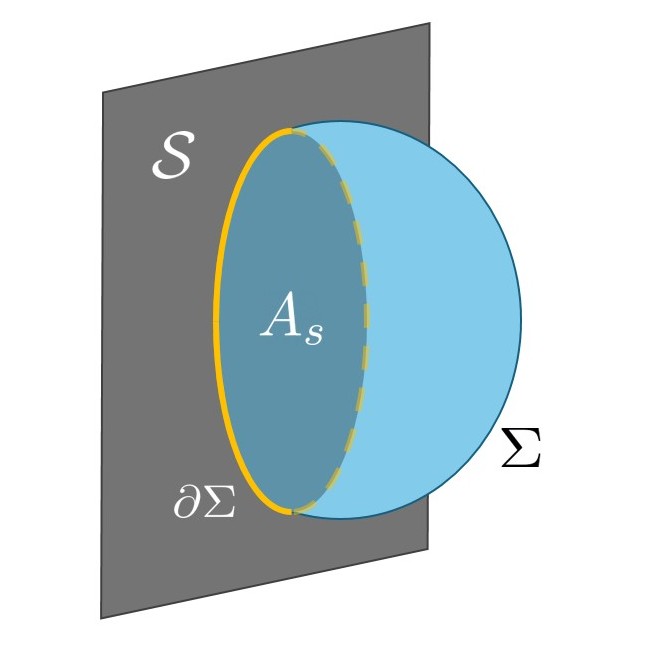}
  \subcaption{A diagram showing the planar bubble-substrate system. The contact hypersurface, $\partial\Sigma$, (yellow) as well as the definition of the area $A_s$.}
  \label{fig:planar-substrate}
\end{subfigure}%
\begin{subfigure}[t]{.5\textwidth}
  \centering
  \includegraphics[width=0.95\linewidth]{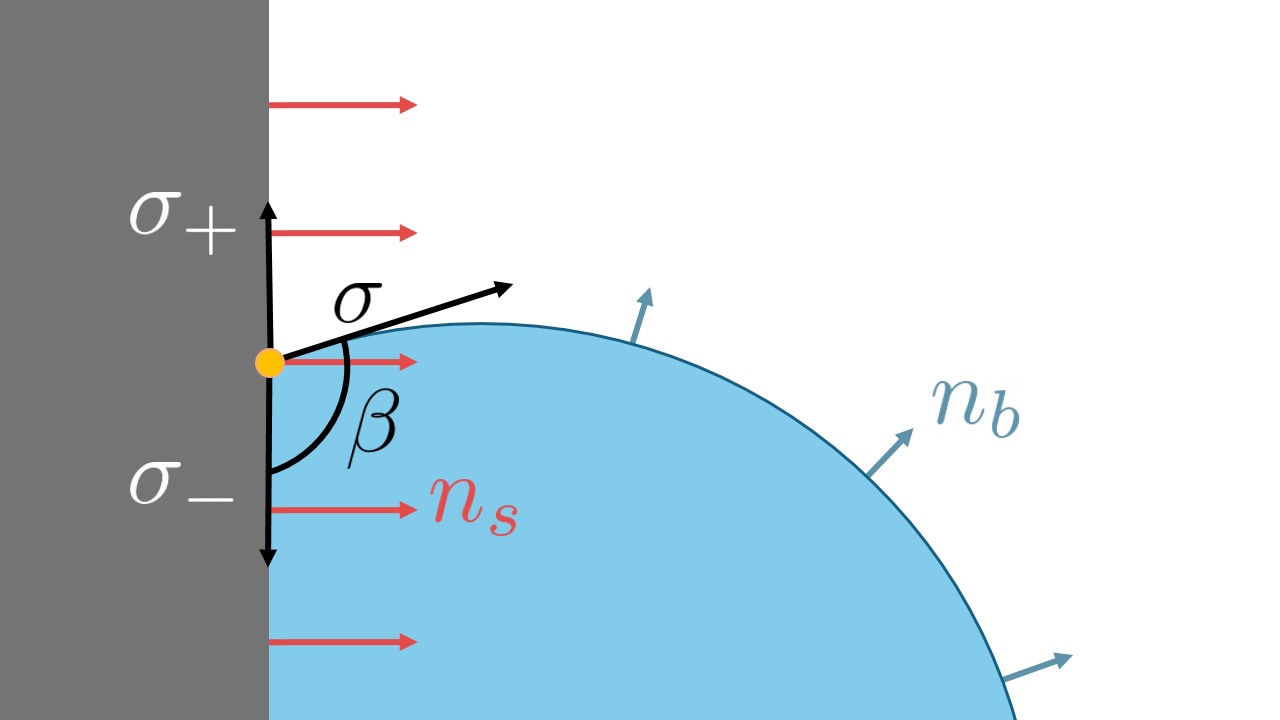}
  \subcaption{A slice of the bubble-substrate system showing the equilibrium junction condition at the contact hypersurface (yellow point).}
  \label{fig:Young-Condition}
\end{subfigure}
\caption{\textbf{(a)} shows the bubble and planar substrate system and the definition of the contact hypersurface $\partial\Sigma$ as well as the substrate, $\mathcal S$ and the bubble hypersurface $\Sigma$. \textbf{(b)} shows how the Dupr\'e-Young condition determines the equilibrium junction condition in Eq.~(\ref{Young-Condition}) for the bubble (blue) nucleating on the substrate (grey). The blue arrows represent the normal vector field to the bubble surface, $n_b$, and the red arrows represent the normal vector to the substrate, $n_s$.}
\label{fig:junction-conditions-bubbles}
\end{figure*}

\item \textit{The Dupr\'e-Young condition}: imposes a junction condition between the surface tensions at the contact hypersurface given by, 
\begin{equation}\label{Young-Condition}
     \cos\beta = \frac{\sigma_- - \sigma_+}{\sigma} = -\frac{\Delta\sigma}{\sigma}.
\end{equation}
$\beta$ is the \textit{equilibrium contact angle} defined as the internal angle between the tangent to the bubble wall and the tangent to the substrate at the contact hypersurface shown in FIG.~\ref{fig:Young-Condition}. 
\end{itemize}

$ $

In light of instanton calculations discussed earlier, the presence of the substrate modifies the Euclidean action in Eq.~(\ref{thin-wall-approximation}), which follows from the discussions of Ref.~\cite{PhysRevD.110.105015}, 
\begin{equation}\label{modified-effective-action}
    \begin{split}
        S_E &= \sigma A -\varepsilon V + \sigma_+A_s - \sigma_-A_s\\
        &=\sigma A -\varepsilon V - \sigma\cos\beta A_s 
    \end{split}
\end{equation}
where $A_s$ is the area on the substrate enclosed by the contact hypersurface, and we have used the Dupr\'e-Young condition to eliminate $\sigma_\pm$.

$ $

For a planar substrate, the area enclosed by the contact hypersurface, $A_s$, in Eq.~(\ref{modified-effective-action}) can be projected onto the bubble surface. We can calculate the areas using a single integration measure $\int \ud A$, where $\ud A$ is the area differential on the bubble surface. We define $\ud A_s$ as the scalar differential area enclosed by the contact hypersurface. These area differentials are related to each other by projecting the components of the vector area differential, $\ud\mathbf{A}$, in the direction of the normal vector to the substrate, ${n}_s$,
\begin{equation}
\begin{split}
    \ud A_s &= {n}_{s}\cdot \ud\mathbf{A}.
\end{split}
\end{equation}
In this way, we can calculate the area enclosed by the contact hypersurface on the bubble,
\begin{equation}
\begin{split} 
    A_s&=\int_\Sigma ({n}_{s}\cdot {n}_b) \ \ud A
\end{split}
\end{equation} 
${n}_b$ is the unit vector field normal to the bubble, and the $\cdot$ represents the inner product between two vectors. Therefore, our action in Eq.~(\ref{modified-effective-action}) can be written more compactly,
\begin{equation}
    \begin{split}
        S_E = -\varepsilon \int \ud V + \sigma \int \ud A\ (1-n_s\cdot n_b \cos\beta) .
    \end{split}
\end{equation} 
This representation of the action will prove useful when we consider the tree-level contributions to the Euclidean action, especially in the case of a bubble forming on a planar substrate.

\section{Non-trivial Field Profiles Around Density Inhomogeneities}
\subsection{Vacuum Screening Inside Dense Spheres}
We now consider some non-trivial solutions of the asymmetron field around a dense spherical object, such as a star \cite{Hinterbichler:2011ca, neutronstars}. In this section, we are primarily interested in obtaining scalar field profiles as $\phi$ undergoes a density-driven phase transition. We will consider the radial step-function density distribution,

\begin{equation}
    \rho(r) = 
        \begin{dcases}
            \rho_s & \text{ for } r<R_s\\
            0 & \text{ for } r>R_s
        \end{dcases}
\end{equation}
where $\rho_s\gg \mu^2M^2$. A spherically symmetric scalar field profile obeys the following \textit{quasi-static} ($\dot\phi=0$) equation of motion, 
\begin{equation}\label{spherically-symmetric}
    \frac{\ud^2\phi}{\ud r^2} +\frac{2}{r}\frac{\ud\phi}{\ud r} = \frac{\ud V_\text{eff}}{\ud\phi}
\end{equation}
with $\phi'(0)=0$ to keep $\phi(r)$ regular at $r=0$ and the asymptotic condition, $\phi(\infty)=\phi_\pm$. We can expand the potential to quadratic order both inside ($r<R_s$) and outside ($r>R_s$) the source, 
\begin{equation}
    V(\phi) = \begin{dcases}
        \frac{\rho_s}{2M^2}\phi^2 & \text{ for } r<R_s\\
        \frac{1}{2}m_\pm^2(\phi-\phi_\pm)^2 & \text{ for } r>R_s.
    \end{dcases}
\end{equation}
where $\phi_\pm$ and $m_\pm$ are defined in Eq.~(\ref{exact-minima}) and Eq.~(\ref{exact-masses}) respectively. We further impose the continuity of $\phi(r)$ and $\phi'(r)$ to ensure $C^1$-smoothness on $\phi(r)$. 

$ $

\begin{figure*}[t]
\includegraphics[width=1\linewidth]{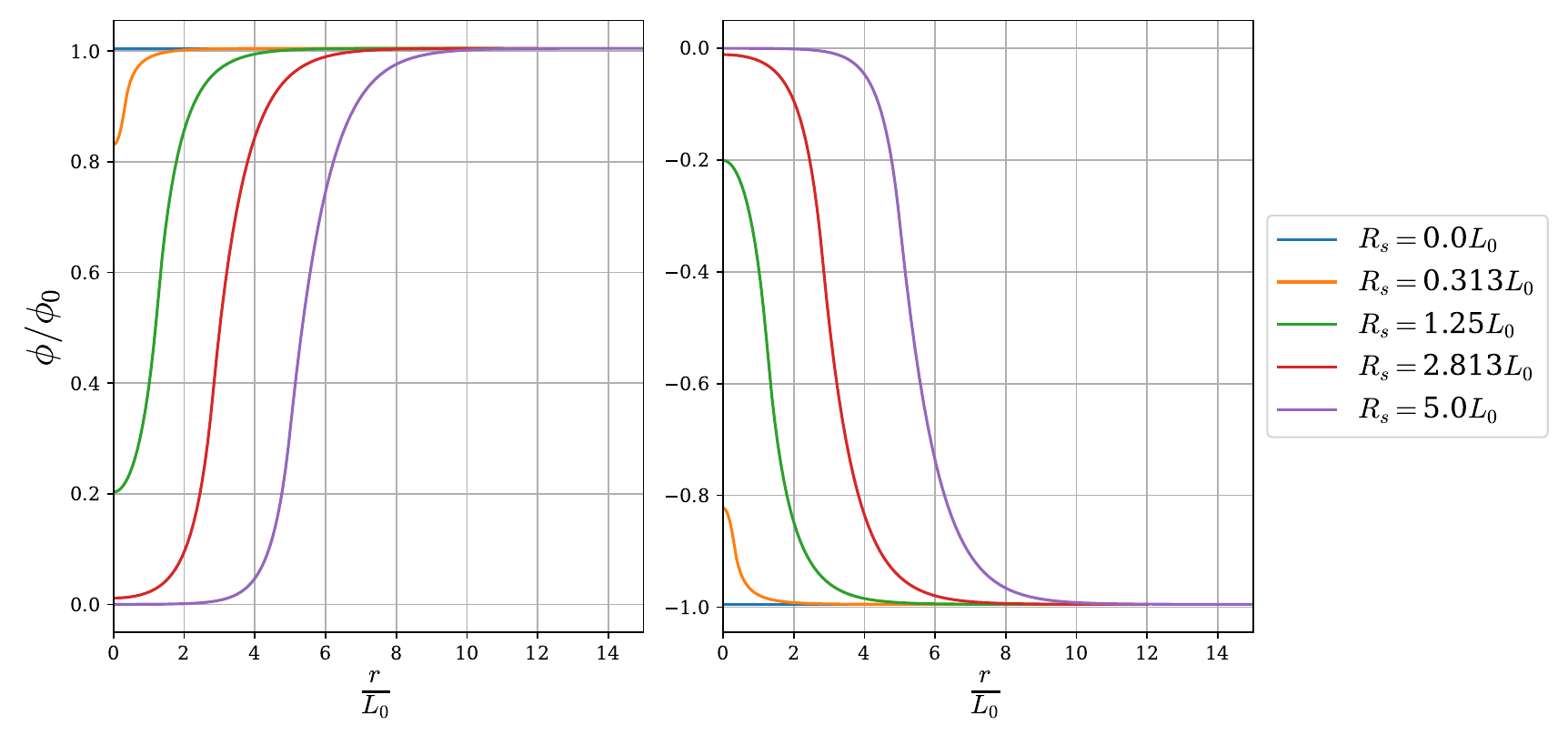}
\caption{Numerical plots of the true (left) and false (right) vacuum field profiles for several values of $R_s$ with $\kappa=0.01\lambda$ and $\rho_s=10\mu^2M^2$. $L_0=(\sqrt{2\mu})^{-1}$ is the symmetron Compton wavelength. The scalar field, $\phi$, is normalised in terms of the magnitude of the VEV of the bare symmetron potential, $\phi_0$. We use the variable $r$ because we are solving the full radial equation found in Eq.~(\ref{spherically-symmetric}).}
\label{fig:vacuum-profile}
\end{figure*}

In Ref.~\cite{Hinterbichler:2011ca}, the scalar field profile around a sphere with arbitrary radius $R_s$ was found. However, we are interested in calculating the energetic properties of the solution in the \textit{thin-shell limit}, $R_s\gg L_\pm$, such that the dynamics of the scalar field are concentrated at the surface of the sphere $r=R_s$. In the fully-planar limit, we further perform a change of variables and  follow Ref.~\cite{udemba2025quantumcorrectionssymmetronfifthforce} to define $z = r-R_s$, in which case, Eq.~(\ref{spherically-symmetric}) is well-approximated by,
\begin{equation}
    \frac{\ud ^2\phi}{\ud z^2} \simeq \frac{\ud V_\text{eff}}{\ud\phi}
\end{equation}
maintaining the boundary conditions imposed on $\phi(r)$ as before. The scalar profile in the thin-shell limit is found to be, 
\begin{equation}\label{asymptotic-solution}
    \phi(z)=\begin{dcases}
        \phi_\text{in}(z) = A_\pm\exp\left(\frac{\sqrt{\rho_s}}{M}z\right) & \text{ for } z<0\\
        \phi_\text{out}(z) = \phi_\pm-B_\pm e^{-m_\pm z} & \text{ for } z>0\\
    \end{dcases}
\end{equation}
where the coefficients $A_\pm$ and $B_\pm$ are determined by matching the solution at $r=R_s$,
\begin{equation}\label{coefficients}
    \begin{split}
        A_\pm &= \frac{\phi_\pm}{1+\frac{\sqrt\rho}{m_\pm M}}, \\ 
        B_\pm &=\frac{\sqrt\rho}{m_\pm M}A_\pm
    \end{split}
\end{equation}
where $\phi_\pm$ and $m_\pm$ are the minima and masses outside the object for $\rho=0$. For $z<0$, the scalar field is screened and is locked at the local minimum $\phi=0$. As the scalar approaches $z=0$, a symmetry-breaking transition occurs, and as $\phi$ leaves the object, it very quickly rolls towards one of the minima of Eq.~(\ref{effective-asymmeton-potential}) as $z\rightarrow \infty$. The variation in $\phi(z)$ is concentrated within a thin shell. The screening mechanism weakly couples $\phi$ to the core of the sphere. Numerical plots of $\phi(z)$ with several values of $R_s$ are shown in FIG.~\ref{fig:vacuum-profile}.

\subsection{Surface Energy Density at the Interface}
In the limit, $R_s\gg L_\pm$, we can calculate the surface energy density of $\phi(z)$. This is given by,
\begin{equation}\label{surface-tension-definition}
\begin{split}
    \sigma &= \int_{-\infty}^\infty \left[\frac{1}{2}\phi'^2 + V(\phi)\right] \ud z\simeq \int_{-c L_0}^{c L_0} \phi'^2 \ud z\\ 
\end{split}
\end{equation}
where $c$ is some $\mathcal{O}(1)$ number greater than unity. In the thin-shell limit, $A_\pm\simeq0$, thus the contributions from inside the sphere are suppressed in Eq.~(\ref{surface-tension-definition}). To the solution $\phi(z)$ in Eq.~(\ref{asymptotic-solution}) approaching $\phi_\pm$ as $z\rightarrow \infty$, we assign the corresponding surface energy defined as $\sigma_\pm$, which is obtained by solving the integral in Eq.~(\ref{surface-tension-definition}), 
\begin{equation}
    \begin{split}
        \sigma_\pm = \frac{m_\pm B_\pm^2}{2} = \frac{\rho_s}{2m_\pm M^2}\frac{\phi_\pm^2}{\left[1+\sqrt{\frac{\rho_s}{m_\pm^2M^2}}\right]^2}\simeq\frac{\phi_\pm^2m_\pm}{2}
    \end{split}.
\end{equation}
The approximation assumes that $R_s\gg L_{\pm}$ and $\rho\gg m_\pm^2M^2$. Through $m_\pm$ and $B_\pm$, $\sigma_\pm$ is a function of $\kappa$, and since we keep $\kappa$ as a small parameter, we capture the leading order behaviour of $\sigma_\pm$ by the expansion given by, 
\begin{equation}
    \sigma_\pm = \sigma_\pm(0) + \kappa\left.\frac{\ud\sigma_\pm}{\ud\kappa}\right|_{\kappa=0} + \mathcal{O}(\kappa^2).
\end{equation}
$\sigma_{\pm}(0)$ is the value of this surface tension in the regular symmetron case, 
\begin{equation}
    \sigma_{\pm}(0) = \frac{\sqrt2 \mu^3}{2\lambda}.
\end{equation}
In the degenerate case ($\kappa=0$), we recover the symmetron potential and $\Delta \sigma=0$.
\begin{figure}
    \centering
    \includegraphics[width=0.8\linewidth]{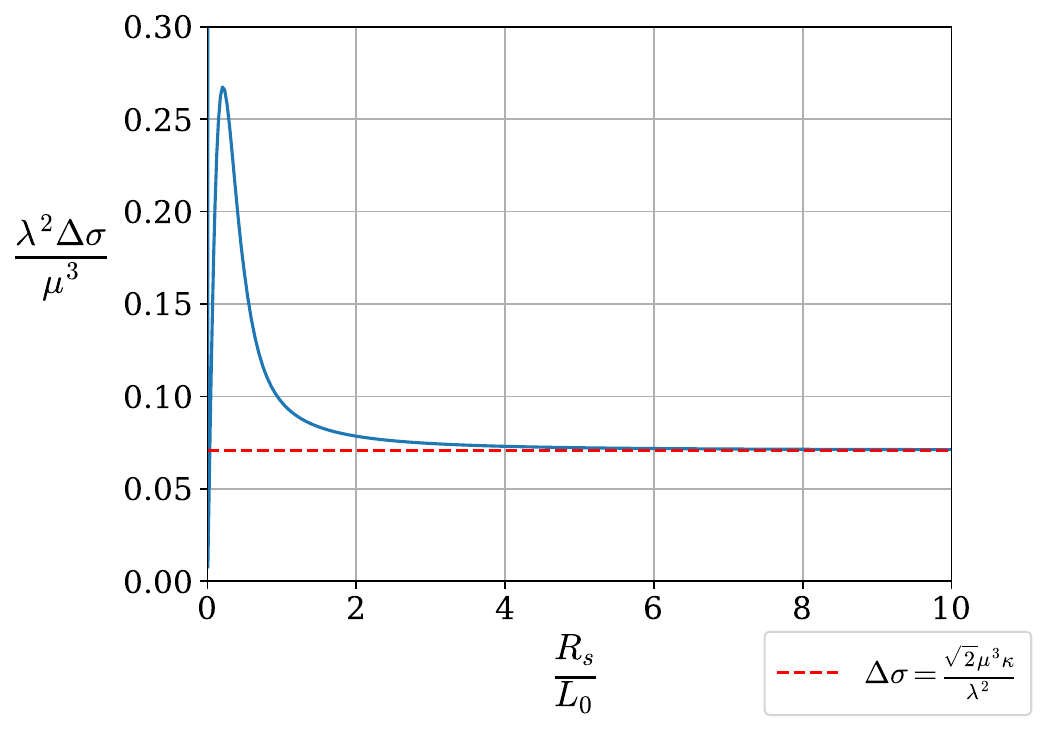}
    \caption{$\Delta\sigma$ versus sphere radius $R_s$ (for $\kappa = 0.01\lambda$ and $\rho_s=10\rho_*$) shows a plateau at large $R_s$. The red-dashed line shows the planar-limit value of $\Delta\sigma$ derived in Eq.~(\ref{junction-condition}).}
    \label{fig:surface-tensions}
\end{figure}
To determine the first-order correction to $\sigma_\pm$, we need to evaluate the first derivative, 
\begin{equation}
\begin{split}
    \left.\frac{\ud\sigma_\pm}{\ud\kappa}\right|_{\kappa=0} &= \left[\frac{\partial\phi_\pm}{\partial\kappa}\frac{\partial\sigma_\pm}{\partial\phi_\pm}\right]_{\kappa=0} + \left[\frac{\partial m_\pm}{\partial\kappa}\frac{\partial\sigma_\pm}{\partial m_\pm}\right]_{\kappa=0} \\
    &= \left[\frac{\partial\phi_\pm}{\partial\kappa}\frac{\partial\sigma_\pm}{\partial\phi_\pm}\right]_{\kappa=0}
\end{split}
\end{equation}
where the second equality is obtained since $m_\pm(\kappa)$ has a local minimum at $\kappa=0$. Thus, the Taylor expansion of $\sigma_\pm(\kappa)$ around $\kappa=0$, is given by, 
\begin{equation}\label{substrate-surface-tension}
    \sigma_\pm(\kappa)\simeq \frac{\sqrt{2}\mu^3}{2\lambda}\left(1\pm\frac{\kappa}{\lambda}\right) + \mathcal{O}(\kappa^2)
\end{equation}
and the difference in the surface tensions, $\Delta \sigma\equiv \sigma_+-\sigma_-$, is non-vanishing, 
\begin{equation}\label{junction-condition}
    \Delta\sigma =\frac{\sqrt{2}\mu^3\kappa}{\lambda^2}.
\end{equation}
This is independent of both $\rho_s$ and $R_s$ as we would expect in the planar limit, since the geometry of the plane is trivial. In FIG.~\ref{fig:surface-tensions}, we see $\Delta\sigma$ approaches a constant as $R_s\rightarrow\infty$. Thus, we see in Eq.~(\ref{Young-Condition}) that the jump in surface tensions, $\Delta\sigma$, is proportional to $\kappa$. This means that we have an angle $\beta = \arccos\left(-\frac{\kappa}{2\lambda}\right)$. For $\kappa\ll\lambda$, this means $\beta = \frac{\pi}{2} + \frac{\kappa}{2\lambda}$ up to higher order corrections.
\begin{figure}[hbt!]
    \centering
    \includegraphics[width=0.75\linewidth]{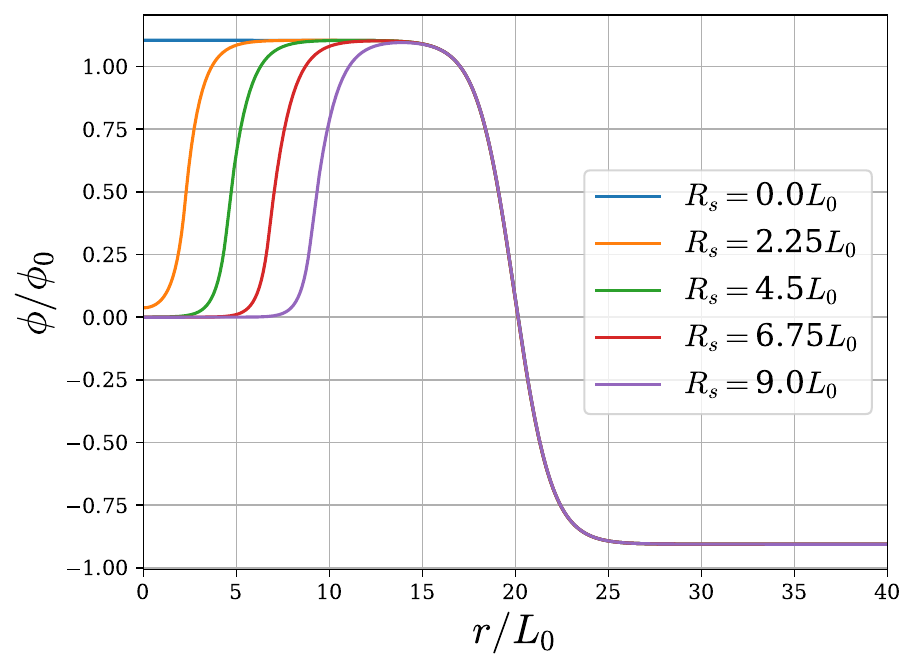}
    \caption{The spherical domain wall coupled to a dense matter sphere with radius $R_s$  (for $\kappa = 0.2\lambda$ and $\rho_s=50\rho_*$).}
    \label{fig:sphaleron-mode-numerical}
\end{figure}
\subsection{Unstable Spherical Domain Walls around Spherical Objects}
Another non-trivial solution of Eq.~(\ref{spherically-symmetric}), with the same boundary conditions, is a spherical domain wall of radius $R$, engulfing a dense spherical object of radius $R_s$. When the domain wall is present, there are additional energy contributions compared to the environment, which come from the surface tension of the domain wall and the $\Delta V$. The numerical profiles of the domain wall are shown in FIG.~\ref{fig:sphaleron-mode-numerical}. Though this solution is difficult to obtain analytically, we can discuss the energy of such a configuration. The total energy of the configuration is given by, 
\begin{equation}\label{sphaleron-mode-energy-functional}
\begin{split}
    E[\phi]&= \int_{\mathbb R^3} \left[\frac{1}{2}(\nabla\phi)^2 + V_\text{eff}(\phi) - V_\text{eff}(\phi_-)\right] \ud\mathcal V.
\end{split}
\end{equation}
Assuming that $R>R_s \gg L_{\pm}$, as arises in the thin-wall approximation, the energy can be expanded in powers of $(L_\pm / R)$ and $(L_\pm / R_s)$,
\begin{equation}\label{sphaleron-mode-energy-functional-thin-wall}
\begin{split}
    E[\phi]& = - \frac{4\pi R^3}{3}\varepsilon + 4\pi R^2\sigma + 4\pi R_s^2\sigma_+\\
    &\ \ \ \ \ -\left(-\frac{4\pi R_s^3}{3}\varepsilon + 4\pi R_s^2\sigma_-\right) + \mathcal{O}\left(\frac{R}{L_\pm^2}\right).
\end{split}
\end{equation}

We have indicated the size of the leading corrections in this expansion, which are linear in the large radial scales; see e.g. Refs.~\cite{IGNATIUS1993252, widyan2012classicalsolutionbouncesecond, Kajantie:1992pn, Gould_2021}. An important consideration in this analysis is the notion of the \textit{critical bubble}. This unstable configuration extremizes the bubble energy, balancing the inward collapse due to the surface tension against the outward pressure difference across the wall. The radius of the critical bubble is given by $R_0 = \frac{2\sigma}{\varepsilon}$. 

$ $

Evaluating the energy on the critical bubble, shown in Fig.~\ref{fig:sphaleron-mode-numerical}, we find,
\begin{equation}
\begin{split}
    E[\phi]& \simeq
    \overbrace{\frac{16\pi\sigma^3}{3\varepsilon^2} + \frac{4\pi\varepsilon}{3} R_s^3}^{\mathcal{O}(m_\pm \kappa^{-2})}
    + \overbrace{4\pi R_s^2\Delta\sigma}^{\mathcal{O}(m_\pm \kappa^{-1})},
\end{split}
\end{equation}
where $\Delta \sigma$ is defined in Eq.~(\ref{junction-condition}), and we have indicated the size of the leading and next-to-leading terms, for simplicity assuming $R_s=\mathcal{O}(R)$. The leading effect of the matter source is to set to zero the contribution to the energy within the source. This decreases the magnitude of the negative energy density contribution from the volume inside the critical bubble in Eq.~(\ref{sphaleron-mode-energy-functional-thin-wall}), thereby suppressing the decay rate for such a configuration. Note that the term containing $\Delta \sigma$ is the same parametric size as terms dropped in Eq.~\eqref{sphaleron-mode-energy-functional-thin-wall}, so to leading order there is no surface energy cost at the boundary of the matter source.

\section{\label{sec:Instantons}The Instantons}
\subsection{Coleman-Callan Instantons coupled to a homogeneous gas}
In the thin-wall approximation, the instanton can geometrically be described as a codimension-1 hypersurface embedded in $\mathbb{R}^{D}$ \cite{PhysRevD.15.2929, PhysRevD.21.3305, Garriga:1994ut, Garriga_1994, CALLAN1998198}. The thin-wall Euclidean action of the instanton offers a low-energy effective description of the scalar dynamics, 
\begin{equation}\label{bulk-thin-wall}
    \begin{split}
        S_E[\phi] \simeq \sigma\int_\Sigma \ud^{D-1}\xi \sqrt{\gamma} - \varepsilon\int_\Omega \ud^D x\sqrt{g}.
    \end{split}
\end{equation}
where $\Sigma=\partial \Omega$ is the bubble hypersurface. The first term is the \textit{Nambu-Goto action} of the bubble wall $\Sigma$ and is proportional to the area of the hypersurface, and $\sigma$ is the surface tension of the bubble in Eq.~(\ref{bounce-profile}). The wall is furnished with coordinates $\xi^a$. The second term is the volume enclosed by the bubble multiplied by $\varepsilon$ from Eq.~(\ref{energy-difference}) across the wall. We will first address the spherical bubbles of Ref.~\cite{PhysRevD.15.2929, PhysRevD.16.1762} but now include corrections from the asymmetron mechanism in the presence of a pressure-free gas of homogeneous matter density $\rho$. 

$ $

To begin with, we write the metric of $\mathbb R^D$ as a foliation of $(D-2)$-spheres, 
\begin{equation}\label{bulk-background-metric}
    \ud s^2 = g_{\mu\nu}\ud x^\mu \ud x^\nu=  \ud\tau^2 + \ud r^2 + r^2\ud\Omega_{D-2}^2
\end{equation}
where $\ud\Omega_n^2$ is the metric over a unit $n$-sphere. The induced metric $\gamma_{ab}$ on the bubble hypersurface, $\Sigma$, is obtained by the pullback map of the background metric onto $\Sigma$, 
\begin{equation}
    \begin{split}
        \gamma_{ab} = \frac{\partial X^\mu}{\partial \xi^a}\frac{\partial X^\nu}{\partial \xi^b}g_{\mu\nu}
    \end{split}
\end{equation}
where $X^{\mu} = X^{\mu}(\xi^a)$ is the function that embeds $\Sigma$ into $\mathbb{R}^D$.
The radial location of the bubble, $r = R(\tau)$, is treated as a free component of the induced metric, $\ud\varsigma^2$, which is written as, 
\begin{equation}
\begin{split}
    \ud\varsigma^2 = \gamma_{ab}\ud\xi^a\ud\xi^b = \left(1+\dot R^2\right)\ud\tau^2 + R^2(\tau)\ud\Omega_{D-2}^2.
\end{split}
\end{equation}
We aim to solve for the function $R(\tau)$. So, the action with the surface and volume terms is evaluated as, 
\begin{equation}\label{bulk-action}
    S_E = 2\varepsilon V_{D-1}\int_{0}^{\tau_m} \ud\tau\left[R_0R^{D-2}\sqrt{1+\dot R^2} - R^{D-1}\right]
\end{equation}
where $R_0=\frac{(D-1)\sigma}{\varepsilon}$ and $V_{D} = \frac{\Omega_{D-1}}{D}$ is the volume enclosed by a unit $D$-ball. $R_0$ is the \textit{critical radius} of the bubble which balances the internal energy contribution and the surface energy such that $S_E$ is minimised for this radius as determined in Ref.~\cite{PhysRevD.15.2929}. This reformulates the instanton calculation into a problem of one-dimensional Lagrangian mechanics, in which we solve for the function, $R(\tau)$. Since the action does not explicitly depend on $\tau$, we have a conserved energy, 
\begin{equation}\label{bubble-energy}
\begin{split}
    E &= p_{_R}\dot R -L=-R^{D-2}\left[\frac{R_0}{\sqrt{1+\dot R^2}} - R\right]
\end{split}
\end{equation}
where $L$ is the Lagrangian of the action in Eq.~(\ref{bulk-action}). On rearranging Eq.~(\ref{bubble-energy}) we obtain the \textit{instanton equation}, which has the form,  
\begin{equation}
    \begin{split}
        \dot R^2 + 2U(R, E)=0
    \end{split}
\end{equation}
where the potential function, $U(R, E)$, is given by,
\begin{equation}
    2U(R, E) = 1-\frac{R_0^2}{[R-ER^{2-D}]^2}.
\end{equation}
We are interested in the $E=0$ case since the instanton calculation assumes a zero input \textit{Euclidean energy}, so we can rearrange to find,
\begin{equation}\label{instanton-solution-bulk}
    \begin{split}
        \tau = \int_{R_0}^{R}\frac{\ud R}{\sqrt{-2U(R)}}
    \end{split}
\end{equation}
where $U(R) = U(R, 0)$. The solution is a $(D-1)$-sphere with radius $R_0$ with the equation given by, $R^2 + \tau^2 = R_0^2$. We can directly calculate the action by using Eq.~(\ref{instanton-solution-bulk}) to eliminate $\tau$ from Eq.~(\ref{bulk-action}),
\begin{equation}
    \begin{split}
        B_0=2\varepsilon V_{D-1}\int _0^{R_0}\ud R \ R^{D-2}\sqrt{R_0^ 2-R^2}
    \end{split}
\end{equation}
where $B_0$ refers to the bulk action in the vacuum ($\rho=0$). A further change of variables to $R=R_0\sin(u)$ gives the following integral expression, 
\begin{equation}\label{reduced-bulk-action}
    B_0= 2\varepsilon V_{D-1} R_0^D\int^\frac{\pi}{2}_0\sin^{D-2}(u)\cos^2(u)\ud u =  \frac{\varepsilon V_{D} R_0^D}{D-1}.
\end{equation}
We are interested in the $D=4$ case for which $B_0$ has the familiar form, 
\begin{equation}\label{d4bulkaction}
    B_0 = \frac{\pi^2\varepsilon}{6}R_0^4 = \frac{27\pi^2\sigma^4}{2\varepsilon^3}
\end{equation}
found in Ref.~\cite{PhysRevD.15.2929}. To obtain the modification due to the symmetron, we promote both $\sigma$ and $\varepsilon$ to their symmetron counterparts, $\sigma\rightarrow \sigma\left(1-\frac{\rho}{\rho_*}\right)^\frac{3}{2}$ and $\varepsilon\rightarrow\varepsilon\left(1-\frac{\rho}{\rho_*}\right)^\frac{3}{2}$,
\begin{equation}\label{bulk-screened-action}
    \begin{split}
        B_\text{bulk} = B_0\left(1-\frac{\rho}{\rho_*}\right)^\frac{3}{2}.
    \end{split}
\end{equation}
The action is suppressed in denser regions and ultimately vanishes at $\rho_*$, since the symmetry of the potential in Eq.~(\ref{effective-potential}) is restored and we only have one minimum at $\phi=0$ --- thus the scalar cannot tunnel as there is no barrier to penetrate. The symmetron modification does not change in other dimensions because the action always obeys the ratio,
\begin{equation}
    B_\text{bulk} \propto\frac{\sigma^D}{\varepsilon^{D-1}}\rightarrow \frac{\sigma^D}{\varepsilon^{D-1}}\left(1-\frac{\rho}{\rho_*}\right)^\frac{3}{2}.
\end{equation}
By following a similar argument, the critical radius, $R_0=\frac{D\sigma}{\varepsilon}$, is unchanged under the modification. 

$ $

Given a sufficiently large static ball of gas with density $\rho$ in Euclidean spacetime, we can assume that the translational invariance of the solutions is approximately preserved, and thus, we retain the form of the decay rate per unit volume, we can compare the decay rate of this system by dividing out by the decay rate of a similar system but at $\rho=0$. The semiclassical contributions to the ratio of decay rates are given by,
\begin{equation}\label{decay_rate-bulk}
    \begin{split}
        \frac{\Gamma(\rho)}{\Gamma(0)} \propto \left(1-\frac{\rho}{\rho_*}\right)^3\exp\left\{{\frac{B_0}{\hbar}\left[1-\left(1-\frac{\rho}{\rho_*}\right)^\frac{3}{2}\right]}\right\}
    \end{split}
\end{equation}
where the proportionality symbol contains the non-zero modes of the fluctuation spectrum. Since zero modes are translational symmetries of the instanton, they constitute a classical contribution to the decay rate, so these contributions are explicitly shown in Eq.~(\ref{decay_rate-bulk}).

$ $

We now compute the decay rate per unit volume, $\Gamma/\mathcal V$, for a bubble forming in a subcritical gas of density $\rho$. We use the BubbleDet package from Ref.~\cite{Ekstedt_2023} to calculate the one-loop functional determinant around the spherically-symmetric background field profile. We shall first define a more simplified representation of $\Gamma/\mathcal V$ that BubbleDet can calculate directly. Firstly, we define the logarithm of the determinant ratio,
\begin{equation}\label{effective-one-loop-action}
    S_1\equiv \frac{1}{2}\ln\left[\frac{\det\mathcal O_b}{\det \mathcal O_-}\right]
\end{equation}
The determinant ratio is defined as the sum of one-loop vacuum Feynman diagrams, including both connected and disconnected ones. The effective one-loop action contains zero modes, negative modes, and positive modes. Thus, the decay rate for a general free bubble can be written as, 
\begin{equation}
    \begin{split}
        \frac{\Gamma}{\mathcal V} = e^{-\frac{(S_0 + S_1)}{\hbar}}
    \end{split}
\end{equation}
where $S_0=B_\text{bulk}$ is the semiclassical contribution to the bulk decay rate. The negative eigenvalue $\ell_{-}$ is also contained in the one-loop effective action, $S_1$. 

$ $

For the numerical calculation using BubbleDet, the ratio is given by, 
\begin{equation}
\begin{split}
    \frac{\Gamma(\rho)}{\Gamma(0)}&=\exp\left[-\frac{\Delta S_0(\rho)+\Delta S_1(\rho)}{\hbar}\right]
\end{split}
\end{equation}
where we defined $\Delta S_i = S_i(\rho) - S_i(0)$.
\begin{figure}[hbt!]
    \centering
    \includegraphics[width=0.75\linewidth]{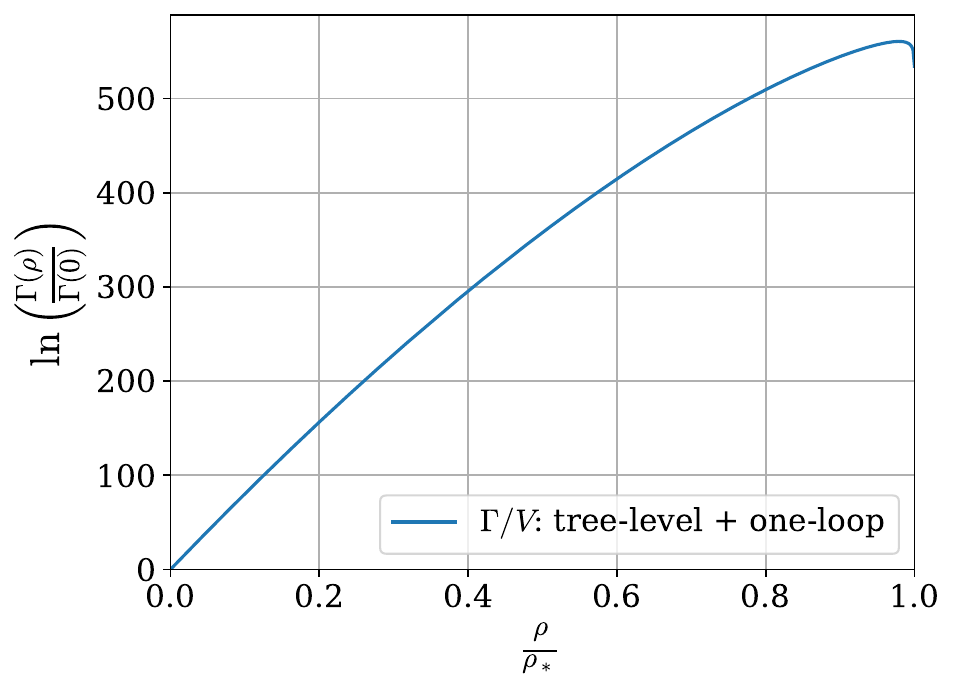}
    \caption{A plot showing the functional dependence of $\frac{\Gamma(\rho)}{\Gamma(0)}$ with $\rho$ ($\kappa=0.9 \lambda$). For $\rho\simeq 0$, $\frac{\Gamma(\rho)}{\Gamma(0)}\sim \mathcal{O}(1)$ and a calculated value of $B_{0} = 547.55\hbar$. For an increasing density, $\frac{\Gamma(\rho)}{\Gamma(0)}$ increases rapidly. As the density is increased beyond $\rho\simeq 0.1\mu^2M^2$, the gradient of the ratio begins to decrease, until a maximum ratio is reached which occurs at a peak density, $\rho_\text{peak}\approx 0.98\rho_*$. Increasing the density beyond $\rho_\text{peak}$ results in a sharp drop-off, suppressing the rate of vacuum bubble production.}
    \label{fig:bubble-in-a-gas}
\end{figure}
In the thin-wall limit, the functional determinant expression can be represented by the geometric properties of the bubble hypersurface as discussed in Ref.~\cite{Garriga:1994ut}. 

$ $

In FIG.~\ref{fig:bubble-in-a-gas}, we show a plot of the ratio of the decay rate as a function of $\rho$. The BubbleDet package in Ref.~\cite{Ekstedt_2023} was used to calculate the ratio of the decay rates since it does not trivially cancel. The functional determinant expression in BubbleDet calculates the sum of all one-loop contributions to the quantum effective action.

$ $

To understand the initial enhancement in the decay rate with density, note that the width of the potential barrier $\Delta \phi$ and the height of the barrier $\Delta V_\text{max}= |V_\text{eff}(\phi_-)|$ both decrease as the density increases. Each of these effects increases the nucleation rate. The sharp decrease as the density approaches $\rho_*$ is due to the zero mode contribution, $\left(1-\frac{\rho}{\rho_*}\right)^3$, dominating over the exponential term. Within the validity of the instanton approximation, this pushes the peak density, $\rho_\text{peak}$, ever-closer to $\rho_*$.  

$ $

Given that the instanton approximation is only valid when $B_\text{bulk}\gg \hbar$, we note that the decay rate is exponentially enhanced for homogeneous densities in the range $0 < \rho \lesssim \rho_*(1-e^{-B_\text{bulk}/(3\hbar)})$, with the upper bound very close to the critical density.
The rate is then suppressed for even higher densities, falling rapidly to zero as $\rho\to\rho_*$ from below.
The peak enhancement occurs at $\rho_\text{peak}\simeq (1-(2\hbar/B_0)^{2/3})\rho_*$, at which point the rate is no longer exponentially suppressed. While the validity of the instanton approximation is compromised at this density, implying that the precise value of $\Gamma(\rho_\text{peak})$ cannot be trusted, the approximation is nevertheless valid either side of this value, and it predicts that the transition proceeds rapidly for $\rho_\text{peak}-\rho\lesssim O((B_0/\hbar)^{-2/3})$, i.e. for densities slightly below the critical density. 
\section{\label{sec:heterogeneous} Nucleation of Bubbles on Curved Surfaces}
\subsection{Bubble Nucleation on Substrates}
\begin{figure*}[t]
    \centering
    \includegraphics[width=0.75\linewidth]{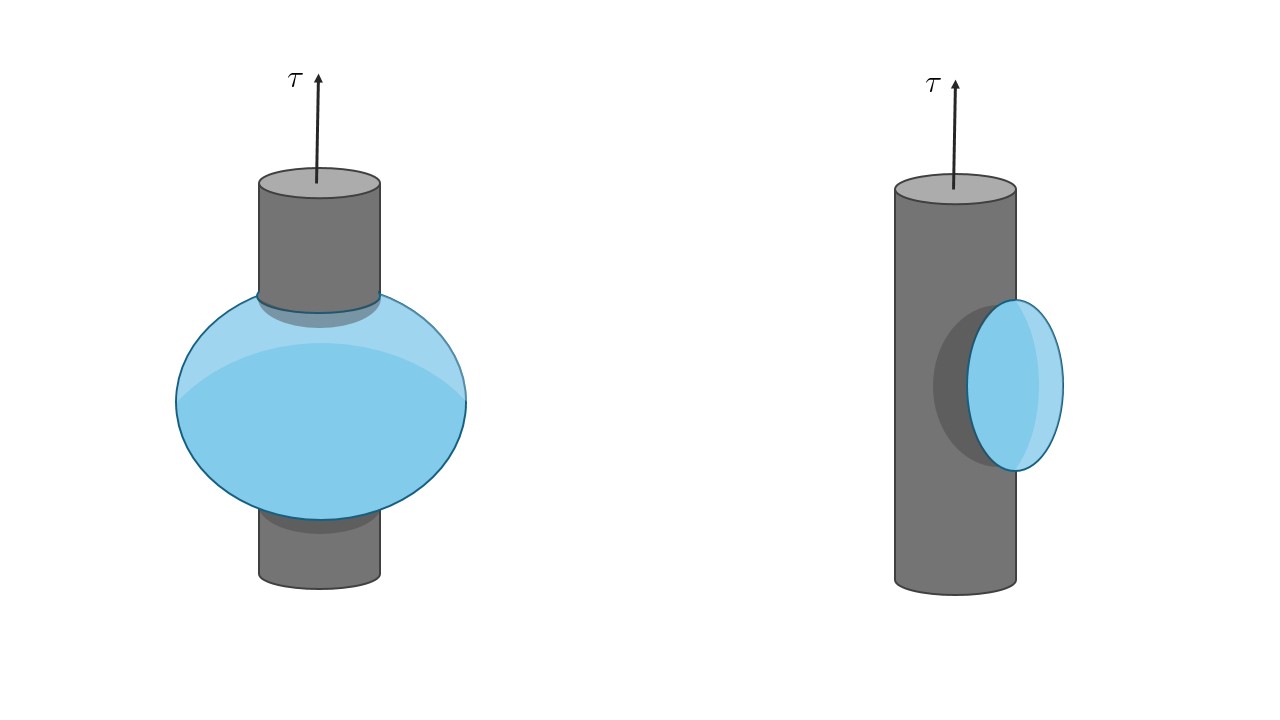}
    \caption{In $\mathbb{R}^D$, bubbles (blue) interact with cylindrical substrates (grey): (left) one fully engulfing the substrate, and (right) one forming at its edge. The time-evolution of the static substrate traces out a cylinder, $\mathbb{R} \times S^{D-2}$.}
    \label{fig:spherical-sub}
\end{figure*}
In the previous analysis, we demonstrated that for $\rho<\rho_*$, the decay rate $\Gamma(\rho)$, for a bubble forming in a gas with density $\rho$, increases with increasing density $\rho$, for larger densities. However, $\Gamma(\rho)$ is maximised for a density $\rho_\text{peak}$, such that $\rho_\text{peak}-\rho_*\ll \rho_*$. When $\rho>\rho_*$, tunnelling can no longer occur, because the barrier of $V_\text{eff}(\phi)$ vanishes. However, another interesting question arises: Can asymmetron bubbles nucleate on the boundary between super- and subcritically dense media? There are several ways in which the bubble can form on such a substrate, as discussed in Ref.~\cite{PhysRevD.110.105015}. We are interested in the following three,
\begin{enumerate}[label=(\roman*)]
    \item Planar Substrate Case - when the bubble forms on the edge of a flat matter source,
    \item Engulfing Cylindrical Case (interstitial in Ref.~\cite{PhysRevD.110.105015}) - when the bubble wraps around a spacetime cylindrical substrate,
    \item Edge Cylindrical Cases - when the bubble forms on the edge of a spacetime cylindrical substrate.
\end{enumerate}
For instanton calculations, we are interested in substrates that are flat in the imaginary time direction; otherwise, this would have the interpretation of a dynamical surface. We have not yet considered the effects or the interpretation of the Wick rotation of a surface that is curved in the imaginary time direction. We also note a point of possible confusion: static spherical objects trace out a cylindrical hypersurface in the imaginary time direction - $\mathcal S = \mathbb R\times S^{D-2}$. So we will refer to such a substrate as \textit{cylindrical}, even though the object itself is a sphere, $S^{D-2}$, in space. For a planar substrate $\mathbb R^{D-2}$, the imaginary time evolution is given by $ \mathcal S = \mathbb R\times \mathbb R^{D-2} = \mathbb R^{D-1}$. Thus, the substrate is the evolution of the static object through imaginary time.

$ $

We further split the edge case into two subcategories: convex edge, when the bubble forms on the exterior of the cylindrical substrate; and the concave edge, when the bubble forms on the interior of a cylindrical substrate. FIG.~\ref{fig:planar-substrate} shows a diagram of a bubble nucleating on a flat planar substrate. FIG.~\ref{fig:spherical-sub} shows diagrams of a bubble nucleating in the engulfing (left-hand figure) and convex (right-hand figure) edge cases. In all cases, we will refer to the curvature radius of the substrate as $R_s$. In what follows, we assume that the substrate has density $\rho>\rho_*$ and is embedded in a matter vacuum, so that bubble nucleation can occur only at or beyond the substrate boundary. Using the modified asymmetron parameters, we will demonstrate how the presence of matter with $\rho<\rho_*$ alters the relevant bubble properties as in Eq.~(\ref{bulk-screened-action}).
\subsubsection{Bubbles on Planar Substrates}
We begin with a most straightforward case of a bubble forming on a planar substrate. The $z$-axis is the spatial direction normal to the plane, which is located at $z=0$ and ${n}_s$ points in the $z>0$ direction. The substrate lies parallel to the mutually-orthogonal axes of imaginary time, $\tau$, and the remaining $D-2$ spatial directions. Thus, the plane is a codimension-1 hypersurface. The normal vector to the bubble must meet the normal of the flat substrate at the contact angle $\beta$, at the contact hypersurface, thus we assume that the system has a symmetry, $SO(D-1)$, for rotations about the $z$-axis. We choose a background metric with this symmetry in mind, 
\begin{equation}\label{planar-background-metric}
    \ud s^2 = \ud z^2 + \ud r^2 + r^2\ud\Omega_{D-2}^2
\end{equation}
where $r$ is the radial displacement from the origin $r$ defined by, $r^2 = \tau^2 + \sum _{i=1}^{D-2}x_i^2$. The function describing the bubble is $r=R(z)$ so the induced metric is, 
\begin{equation}
    \ud\varsigma^2 = \gamma_{ab}\ud\xi^a\ud\xi^b =\left(1+R'^2\right)\ud z^2+R^2(z)\ud\Omega_{D-2}^2.
\end{equation}
The normal vectors on the substrate and the bubble are 
\begin{equation}
    \begin{split}
        {n}_s =\ud z,&\ \ \ {n}_b = \frac{\ud r-R'\ud z}{\sqrt{1+R'^2}}\\
        {n}_b\cdot {n}_s &= -\frac{R'}{\sqrt{1+R'^2}}.
\end{split}
\end{equation}
Thus, we construct the Euclidean action, 
\begin{equation}\label{flatsubstrateaction}
    \begin{split}
        S_E = \varepsilon V_{D-1}\int_0^{z_m} \ud z\ R^{D-2}\left[R_0\sqrt{1+R'^2} + R_0 R'\cos\beta-R\right]
    \end{split}
\end{equation}
which is structurally similar to the integrand in Eq.~(\ref{bulk-action}). Since the Lagrangian does not depend explicitly on $z$, we can derive a conserved quantity once more given by
\begin{equation}
    \Pi \equiv -R^{D-2}\left[\frac{R_0}{\sqrt{1+R'^2}}-R\right]=0.
\end{equation}
$\Pi$ and is called the spatial momentum density and is identical in structure to the expression for $E$ in Eq.~(\ref{bubble-energy}), and the analysis of the previous calculation follows. Again, we set $\Pi=0$ and find that the bubble profile is described by a spherical cap. We use the Dupr\'e-Young condition to obtain a natural boundary condition for the bubble at the substrate, given by $R'(0) = \tan\left(\frac{\pi}{2}-\beta\right)$, (where $'$ denotes a derivative with respect to $z$), so we find the spherical cap is parametrised by $\beta$,
\begin{equation}\label{flat-bubble-surface}
    \begin{split}
        R(z)= \sqrt{R_0^2-(z+R_0\cos\beta)^2} \ \ \ \text{ for } z>0
    \end{split}
\end{equation}
which is the equation of a spherical cap with the correct equilibrium contact angle on the substrate. The integration limit in Eq.~(\ref{flatsubstrateaction}) is given by, 
\begin{equation}
    z_m = R_0(1-\cos\beta).
\end{equation}
$B$ is obtained by substituting the instanton equation and changing to an angular variable, $R =\pm R_0\sin(u)$,
\begin{widetext}
\begin{equation}
    \begin{split}
    B_\text{p}(\beta) =\varepsilon V_{D-1} R_0^{D}\int_{0}^{\beta}\sin^{D-2}(u)\left[\cos^2(u)-\cos(\beta)\cos(u)\right]\ud u.
    \end{split}
\end{equation}
\end{widetext}
The integral here deviates from Eq.~(\ref{reduced-bulk-action}), with the additional boundary substrate term --- this additional term subtracts the volume inside the substrate from the volume enclosed by the spherical bubble to give the volume of the spherical cap. For example, in the $D=4$ case, the action is given by
\begin{equation}\label{planard4}
    \begin{split}
        B_\text{p}(\beta) &=\frac{B_0}{\pi}\left[\beta-\sin\beta\cos\beta - \frac{2}{3}\sin^3\beta\cos\beta\right]
    \end{split}
\end{equation}
as found in Ref.~\cite{PhysRevD.110.105015} with $B_0$ defined in Eq.~(\ref{d4bulkaction}). Since $\beta\simeq\frac{\pi}{2}$, the bubble is approximately a hemisphere with action, $B_\text{p}(\frac{\pi}{2}, 0) \simeq \frac{1}{2}B_0$. When the density outside the substrate is $0<\rho<\rho_*$, we can promote $\sigma$ and $\varepsilon$ to their symmetron counterparts to obtain the density dependence of the planar substrate-bubble system, 
\begin{widetext}
\begin{equation}
    \begin{split}
        B_\text{p}(\beta, \rho) = \frac{B_0}{\pi}\left[\beta-\sin\beta\cos\beta - \frac{2}{3}\sin^3\beta\cos\beta\right]\left(1-\frac{\rho}{\rho_*}\right)^\frac{3}{2}
    \end{split}
\end{equation}
\end{widetext}
The symmetron modification has the same effect in this case as the bubble nucleating in the bulk of the scalar field.

$ $

The decay rate for bubble nucleation on such a substrate in $D=4$, is given by, 
\begin{equation}
    \begin{split}
         \frac{\Gamma_\text{p}}{\mathcal A} = \left(\frac{B_\text{p}}{2\pi\hbar}\right)^{\frac{3}{2}}\frac{1}{\sqrt{|\ell_-|}} \left(\frac{\overline{\det}[-\partial_\mu\partial_\mu+V_\text{eff}''(\phi_b)]}{\det[-\partial_\mu\partial_\mu+V_\text{eff}''(\phi_-)]}\right)^{-\frac{1}{2}}e^{-\frac{B_\text{p}}{\hbar}}.
    \end{split}
\end{equation}
Since the plane breaks translational invariance of the bounce in the $z$ direction, we lose one zero mode in the spectrum, and thus, the method of collective coordinates \cite{PhysRevD.110.116023, Paranjape:2017fsy, Coleman:1985rnk} results in a factor of the area, $\mathcal A$, instead of the conventional volume factor, $\mathcal V$. Essentially, a particular bubble with some radius can form anywhere on the plane with a contact angle $\beta$, but cannot form arbitrarily far away from the plane. The loss of this zero mode is also expressed in the $\left(\frac{B_\text{p}}{2\pi\hbar}\right)^\frac{3}{2}$ factor, which can be compared to the decay rate initially described Eq.~(\ref{decay_rate-bulk}). The negative eigenvalue is denoted by $\ell_-$.
\subsubsection{Engulfing Instanton}
For cylindrical substrates, we will first consider the case of the engulfing instanton. In this case, a spacelike slice of the diagram in FIG.~\ref{fig:spherical-sub} through the bubble results in a pair of concentric $(D-1)$-spheres, the smaller one representing the substrate and the larger one the bubble wall. The time evolution of the bubble is given by a function $r=R(\tau)$, and the background metric is the same as Eq.~(\ref{bulk-background-metric}).

$ $ 

\begin{figure}
  \centering
  \includegraphics[width=0.75\linewidth]{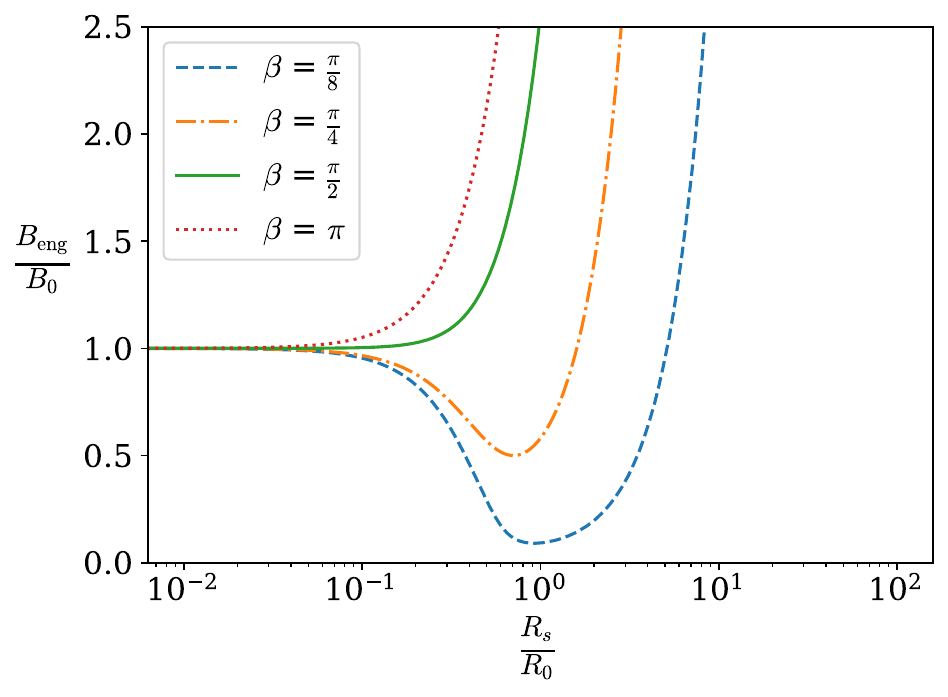}
\caption{Numerical plots showing how the action in Eq.~(\ref{engulfing-instanton}), $B_\text{eng}$, of an engulfing bubble varies with the seed radius, $R_s$, in $D=4$ with various contact angles: $\beta=\frac{\pi}{8}, \frac{\pi}{4}, \frac{\pi}{2}, \pi$.}
\label{fig:action-engulfing}
\end{figure}

The background metric, $\ud s^2$, and induced metric, $\ud\varsigma^2$, are given by, 
\begin{equation}
    \begin{split}
        \ud s^2 &= \ud\tau^2 + \ud r^2 + r^2\ud\Omega_{D-2}^2\\
        \ud\varsigma^2 &=\left(1+\dot R^2\right)\ud\tau^2 + R^2(\tau)\ud\Omega_{D-2}^2.
    \end{split}
\end{equation}
Thus, the action for this system is given by, 
\begin{widetext}
\begin{equation}
    \begin{split}
        S_E &= 2\varepsilon V_{D-1}\int_0^{\tau_s}\ud\tau\left[R^{D-2}\left(R_0\sqrt{1+\dot R^2}-R\right)-R_s^{D-2}\left(R_0\cos\beta-R_s\right)\right].
    \end{split}
\end{equation}
\end{widetext}
The conserved energy $E$ is given by,
\begin{equation}
    E = -\frac{R_0 R^{D-2}}{\sqrt{1+\dot R^2}} + R^{D-1} = -R_0R_s^{D-2}\cos\beta + R_s^{D-1}. 
\end{equation}
Finally, we change variables using the instanton equation to obtain, 
\begin{equation}\label{engulfing-instanton}
    B_\text{eng} = 2\varepsilon V_{D-1}\int_{R_s}^{R_*} \ud R R^{D-2}\sqrt{R_0^2-(R-ER^{2-D})^2}.
\end{equation}
where $R_*$ is the largest positive root of the equation,
\begin{equation}\label{largest-root}
    R_*^{D-1} \pm R_0 R_*^{D-2} - E = 0.
\end{equation}
and $R_*$ depends on $R_s$ through Eq.~(\ref{largest-root}). The solution to the Euclidean action $B_\text{eng}$ is numerically calculated for several angles for spheres of various radii, $R_s$ and presented in FIG.~\ref{fig:action-engulfing}. For $\beta\geq\frac{\pi}{2}$, we see that for $R_s\ll R_0$, then the Euclidean action is close to that of the bulk bubble. As the seed is made bigger, a larger critical bubble must be nucleated, which has a greater Euclidean action cost, making the bubble less likely to engulf a large seed. This motivates edge nucleation in Ref.~\cite{PhysRevD.110.105015}.

$ $

For angles $\beta<\frac{\pi}{2}$, the small $R_s$ behaviour is much the same. However, when $R_s\sim R_0$, the Euclidean action acquires a local minimum, for which the rate of bubble nucleation is enhanced. For $R_s\gg R_0$, the Euclidean action again increases as a larger bubble must be formed, which again suppresses the rate of bubble nucleation on larger seeds. 

$ $

While $\beta>\frac{\pi}{2}$ exhibits interesting behaviour for $R_s\sim R_0$, the only physically relevant range of angles for the nucleation of asymmetron bubbles satisfies the inequality, $\beta\gtrsim \frac{\pi}{2}$.
Thus, for the angles we are interested in, the Euclidean action is always greater than $B_0$.
\subsubsection{Edge Instantons: Perturbative Expansion}
\label{sec:perturbative}
\begin{figure*}
    \centering
    \includegraphics[width=0.75\linewidth]{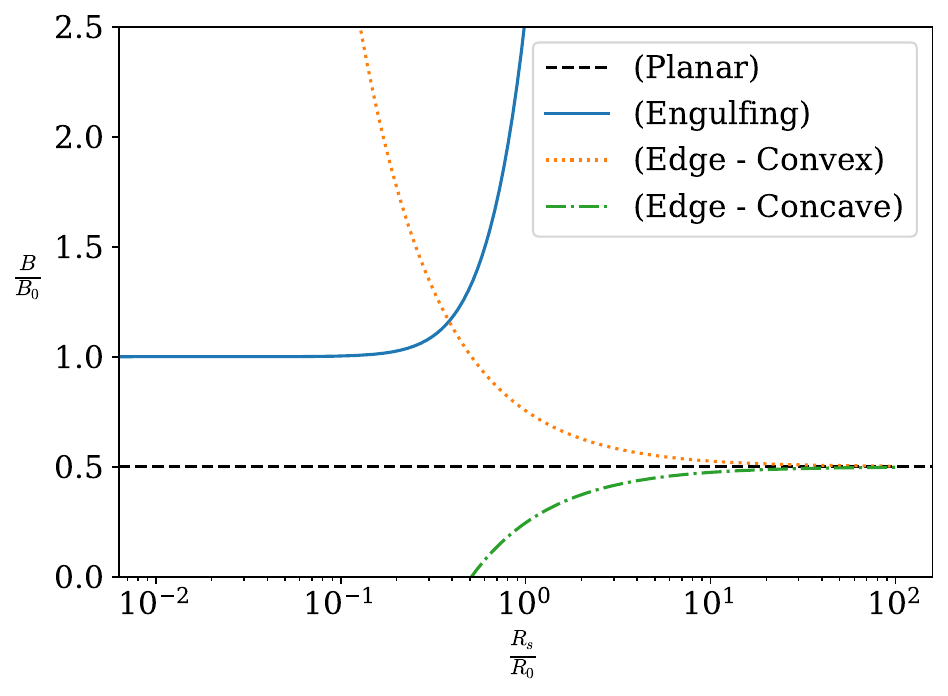}
    \caption{$B$ for $D=4$ instantons forming on cylindrical substrates with curvature radius $R_s$ with $\beta=\frac{\pi}{2}$. Additionally, the action of the bubble forming on a flat, planar substrate is plotted. It can be seen that the edge spherical actions approach the flat planar action in the large $R_s$ limit.}
    \label{fig:spherical-actions}
\end{figure*}

To extend the analysis to nucleation of bubbles on a cylindrical substrate, we can perturbatively calculate corrections to the Euclidean action from the planar case introduced in Ref.~\cite{Soleimani}. Thus, the Euclidean action can be split into the planar contribution and an $R_s$-dependent perturbation part, 
\begin{equation}\label{correction-euclidean-action}
    B_\text{edge} = B_\text{p} + \Delta B
\end{equation}
where $\Delta B$ is the $\mathcal O(R_s^{-1})$ correction to the Euclidean action. For simplicity, we consider the bubble forming on the edge of a cylindrical substrate. The correction to the Euclidean action in Eq.~(\ref{correction-euclidean-action}) on a convex (taking the $+$ sign) or concave (taking the $-$ sign) cylindrical surface is given by,
\begin{equation}\label{perturbed-spherical-action}
    \Delta B = \pm\frac{2\pi R_0^4\varepsilon}{15}\left(\frac{R_0}{R_s}\right)\sin^5\beta.
\end{equation}
The term with the $+$ sign was determined in Ref.~\cite{PhysRevD.110.105015}. The $-$ sign is explained by following the conventions in Ref.~\cite{Soleimani}, or rather more simply using the calculation presented in Ref.~\cite{PhysRevD.110.105015}, and switching the sign in terms containing $R_s$. It can be seen that the smallest Euclidean action is achieved by a bubble forming on a concave substrate (such as the interior of a spherical cavity or vacuum chamber), which takes the negative sign in Eq.~(\ref{perturbed-spherical-action}). The concave and convex cylindrical substrate actions are plotted in FIG.~\ref{fig:spherical-actions} for $\beta=\frac{\pi}{2}$ along with the actions for the corresponding planar and engulfing bubbles. The edge cases diverge for $R_s\ll R_0$ as the perturbative calculation breaks down. 

$ $

\section{\label{sec:discussion}Discussion}
\subsection{What is the preferred nucleation channel?}
Consider an approximately spherical vacuum chamber with radius $R_s$ and the density inside the sphere $\rho=0$. The instanton equations are solved by the slightly concave solution or the spherical bulk instanton with $\rho=0$ in Eq.~(\ref{bulk-action}). We can compare the decay rates to determine whether there is a preference for a particular bubble at some radius of curvature. We know that the concave edge case has the smallest Euclidean action, making it the preferred decay channel semiclassically. We are interested in whether quantum corrections can present a maximum curvature radius for which edge nucleation is preferred. The decay rates for the edge and bulk cases are given by,
\begin{widetext}
\begin{equation}
    \begin{split}
        \Gamma_\text{edge} = 4\pi R_s^2 \times\left(\frac{B_\text{edge}}{2\pi\hbar}\right)^{\frac{3}{2}}\times\frac{1}{\sqrt{|\ell^\text{edge}_{-}|}}\times \left(\frac{\overline{\det}[\mathcal O_\text{edge}]}{\det[\mathcal O_{\text{edge},-}]}\right)^{-\frac{1}{2}}e^{-B_\text{edge}/\hbar}\\
        \Gamma_\text{bulk} = \frac{4}{3}\pi R_s^3 \times\left(\frac{B_\text{bulk}}{2\pi\hbar}\right)^{2}\times\frac{1}{\sqrt{|\ell^\text{bulk}_{-}|}}\times \left(\frac{\overline{\det}[\mathcal O_\text{bulk}]}{\det[\mathcal O_{\text{bulk},-}]}\right)^{-\frac{1}{2}}e^{-B_\text{bulk}/\hbar}\\   
    \end{split}
\end{equation}
\end{widetext}
where $\overline{\det}[\mathcal O]$ is the determinant of the fluctuation spectrum with zero and negative modes removed. We take the ratio of the decay rates, $\Gamma_\text{edge}/\Gamma_\text{bulk}=\mathcal K$, which is given by,
\begin{equation}
    \begin{split}
        \mathcal{K} =  \frac{3}{ R_s}\times\frac{\left(\frac{B_\text{edge}}{2\pi\hbar}\right)^{\frac{3}{2}}}{\left(\frac{B_\text{bulk}}{2\pi\hbar}\right)^{2}}\times\sqrt{\frac{|\ell^\text{bulk}_{-}|}{|\ell^\text{edge}_{-}|}}\times\frac{\left(\frac{\overline{\det}[\mathcal O_\text{edge}]}{\det[\mathcal O_{\text{edge},-}]}\right)^{-\frac{1}{2}}}{ \left(\frac{\overline{\det}[\mathcal O_\text{bulk}]}{\det[\mathcal O_{\text{bulk},-}]}\right)^{-\frac{1}{2}}}\times e^{\Delta B/\hbar}
    \end{split}
\end{equation}

where $\Delta B = B_\text{bulk}-B_\text{edge}$. It is not immediately obvious that this is dimensionless, since we have an overall dependence on the radius of the vacuum chamber $R_s$. However, heuristically, we know that the full ratio of determinants is dimensionless, 
\begin{equation}
    \begin{split}
        \left(\frac{\det[\mathcal O_\text{bulk}]}{\det[\mathcal O_{\text{bulk},-}]}\right)^{-\frac{1}{2}}.
    \end{split}
\end{equation}
The zero modes make this expression formally divergent; thus, one employs the method of integrating over collective coordinates by performing an integration over the location of the instantons \cite{PhysRevD.110.116023}. The usual method would turn the above expression into one that looks like this, 
\begin{widetext}
\begin{equation}
    \begin{split}
        \left(\frac{\det[\mathcal O_\text{bulk}]}{\det[\mathcal O_{\text{bulk},-}]}\right)^{-\frac{1}{2}}\xrightarrow[\text{removed}]{\text{4 zero modes}} T\times\mathcal V_\text{chamber}\times \left(\frac{B_\text{bulk}}{2\pi\hbar}\right)^2\times\left(\frac{\widetilde\det[\mathcal O_\text{bulk}]}{\det[\mathcal O_{\text{bulk},-}]}\right)^{-\frac{1}{2}} 
    \end{split}
\end{equation}
\end{widetext}
where $\widetilde{\det}$ is the fluctuation determinant with the zero modes removed. This means that the functional determinant ratio on the right-hand side has an overall dimensionality of $[\text{length}]^{-4}$. Similarly, for the edge case, the method of collective coordinate would result in the following expression,
\begin{widetext}
\begin{equation}
    \begin{split}
        \left(\frac{\det[\mathcal O_\text{edge}]}{\det[\mathcal O_{\text{edge},-}]}\right)^{-\frac{1}{2}}\xrightarrow[\text{removed}]{\text{3 zero modes}} T\times\mathcal A_\text{chamber}\times \left(\frac{B_\text{edge}}{2\pi\hbar}\right)^{\frac{3}{2}}\times\left(\frac{\widetilde\det[\mathcal O_\text{edge}]}{\det[\mathcal O_{\text{edge},-}]}\right)^{-\frac{1}{2}}.
    \end{split}
\end{equation}
\end{widetext}
In this case, the functional determinant on the right-hand side has an overall dimensionality of $[\text{length}]^{-3}$. 

$ $

\begin{figure}
    \centering
    \includegraphics[width=0.75\linewidth]{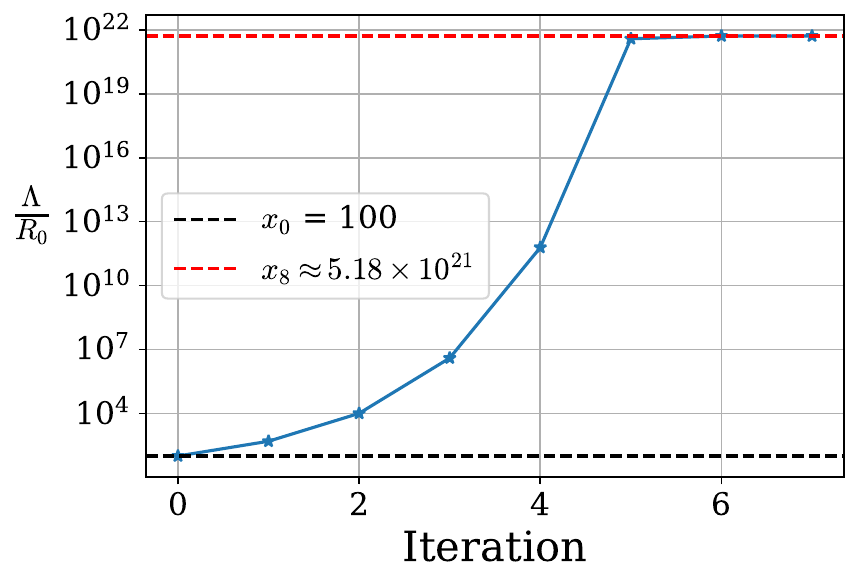}
    \caption{A graph showing the convergence of the solution of $\Lambda$ to Eq.~(\ref{sign-change}) using the Newton-Raphson method with a representative values of $B_0=100\hbar$, $\rho=0$ and $\beta=\frac{\pi}{2}$ so that $f(\beta)=\frac{1}{2}$. We use a tolerance of $\delta x=10^{-10}$, and we observe convergence after only $8$ iterations of the Newton-Raphson scheme. It can be seen that even for small values of $B_0$ within the instanton approximation, this critical radius converges on an enormous factor of $R_0$, $\Lambda\simeq 10^{22}R_0$, which matches the order-of-magnitude estimate in Eq.~(\ref{rough-sketch-exponentiation}).}
    \label{fig:newton-raphson}
\end{figure}

Assuming both fluctuation operators only contain one negative mode each, this means that there is one additional positive fluctuation mode in the edge case operator compared to the bulk case, which we need to isolate from the fluctuation spectrum to explicitly show the correct dimensions in the formula for the ratio. This gives the following expression for the ratio $\mathcal K$,
\begin{equation}
    \begin{split}
        \mathcal{K} \propto  \frac{R_0}{ R_s}\times e^{\Delta B/\hbar}
    \end{split}
\end{equation}
We wish to condense all of the length dependence into the exponential part of the expression without introducing further length scales into the problem. Irrespective of the nuance of evaluating the functional determinants and negative eigenvalues in the thin-wall approximation, the only length scale available to the instanton will be $R_0$; thus, at least heuristically, the negative modes of both the bulk and edge cases are $\propto R_0^{-2}$. This is explicitly shown for the bulk instanton in Ref.~\cite{Paranjape:2017fsy}. The ratio of the negative modes does not contribute any new length scale to the formula. Computing the one-loop functional determinant as done in Ref.~\cite{Ekstedt_2023}, the standard result is that the tree-level contributions are scaled by the exponential of a dimensionless number, $e^{K}$, where $K$ is an $\mathcal{O}(1)$ number. When combined with the semiclassical exponent, $K$ is much smaller by several orders of magnitude than the Euclidean action; thus, we shall ignore it in the subsequent analysis. Thus, we obtain the effective action difference, $\mathcal{J}_\text{eff}$, 
\begin{equation}
    \mathcal{J}_\text{eff} = \frac{\Delta B}{\hbar} - \ln\left(\frac{R_s}{R_0}\right)
\end{equation}
where $\Delta B$ is defined as, 
\begin{equation}
    \begin{split}
        \Delta B = B_\text{bulk} \left[1-f(\beta)+ \frac{4}{5\pi}\left(\frac{R_0}{R_s}\right)\sin^5\beta\right]
    \end{split}
\end{equation}
where $f(\beta) = \frac{1}{\pi}\left[\beta-\sin\beta\cos\beta-\frac{2}{3}\sin^3\beta\cos\beta\right]$. The exponent is then given by, 
\begin{equation}
    \mathcal{J}_\text{eff} = \frac{B_\text{bulk}}{\hbar} \left[1-f(\beta)\pm \frac{4}{5\pi}\left(\frac{R_0}{R_s}\right)\sin^5\beta\right] - \ln\left(\frac{R_s}{R_0}\right)
\end{equation}
\begin{itemize}
    \item If $\mathcal{J}_\text{eff}>0$ then the edge case is the dominant decay process,
    \item but, if $\mathcal{J}_\text{eff}<0$, then the bulk case is the dominant decay process. 
\end{itemize}
But the equation for $\mathcal{J}_\text{eff}=0$ is transcendental, so an explicit expression that can be numerically solved using the Newton-Raphson scheme as shown in FIG.~\ref{fig:newton-raphson}. In any case, we can obtain a leading-order closed-form expression within the regime of our approximations, 
\begin{equation}\label{sign-change}
\begin{split}
    \frac{R_s}{R_0} &= e^{\frac{B_\text{bulk}}{\hbar}[1-f(\beta)]}\times\exp\left[\frac{4B_\text{bulk}}{5\pi\hbar}\left(\frac{R_0}{R_s}\right)\sin^5\beta\right].\\ 
\end{split}
\end{equation}
The convergence of the Newton-Raphson scheme is shown in FIG.~\ref{fig:newton-raphson}, using a representative value of $B_0=100 \hbar$, $\rho=0$ and $\beta=\frac{\pi}{2}$ so that $f(\beta)=\frac{1}{2}$. We solve the equation, 
\begin{equation}
    h(x) = x - e^{\frac{B_\text{bulk}}{2\hbar}}\times\exp\left[\frac{4B_\text{bulk}}{5\pi\hbar}\left(\frac{1}{x}\right)\right]=0.
\end{equation}
where $x = \frac{R_s}{R_0}$. The iterated values of the Newton-Raphson scheme are calculated by the formula, 
\begin{equation}
    x_{n+1} = x_n-\frac{h(x_n)}{h'(x_n)},
\end{equation}
where we have used $x_0 = 100$, we define convergence by a tolerance such that when $\delta x = x_{n+1}-x_n=10^{-10}$, the iterative scheme is terminated.

$ $

On the contrary, the scale hierarchy of the thin-wall and perturbative approximations used to obtain the action of the instanton forming on a concave substrate: $L_0\ll R_0\ll R_s$. The first inequality is the regime of the thin-wall approximation, and the second inequality gives the perturbative regime for the approximately planar substrate. We can then expand the expression in terms of powers of $\left(\frac{R_0}{R_s}\right)$ such that the leading-order contribution to $R_s/R_0$ is given by,
\begin{equation}
    \Lambda \simeq R_0e^{\frac{B_\text{bulk}}{\hbar}[1-f(\beta)]}
\end{equation}
where $\Lambda$ is the \textit{critical radius of curvature} of the chamber for which $\mathcal J_\text{eff}\simeq0$. In the case of a hemispherical bubble $\beta=\frac{\pi}{2}$, this critical radius for a substrate surrounded by a matter vacuum is approximately, 
\begin{equation}\label{hemispherical-limit}
    \Lambda \simeq R_0e^{\frac{B_\text{0}}{2\hbar}}.
\end{equation}
Thus, if $R_s<\Lambda$, then the decay of the false vacuum on the edge is dominant, whereas for $R_s>\Lambda$, the bulk decay is dominant. Using the representative value presented in FIG.~\ref{fig:newton-raphson}, we see that the result of this approximation gives an estimate for the order of magnitude of the ratio $\Lambda/R_0$, since 
\begin{equation}\label{rough-sketch-exponentiation}
    \frac{\Lambda}{R_0}\simeq e^{50}\simeq 10^{21.7}.
\end{equation}
The instanton calculation is valid when $B_\text{0}\gg\hbar$ and thus, the proportionality constant between $\Lambda$ and $R_0$ is a large number, so this critical radius $\Lambda$ lies in the region of validity of the perturbative spherical edge calculations performed earlier.
\subsection{Estimating $\Lambda$ in Vacuum Chamber Experiments}
Using the estimated values of the parameters of the symmetron suggested in Ref.~\cite{Clements:2023bva}, we can provide an estimate of this upper limit of edge nucleation in a vacuum chamber with radius $R_s$. Then we can apply the symmetron modification to Eq.~(\ref{hemispherical-limit}) and obtain, 
\begin{equation}\label{upper-limit-screened}
    \begin{split}
        \Lambda  = R_0\exp\left[\frac{B_0}{2\hbar}\left(1-\frac{\rho}{\rho_*}\right)^{\frac{3}{2}}\right].
    \end{split}
\end{equation}
$B_\text{0}$ is given by Eq.~(\ref{d4bulkaction}) and we use a representation of $R_0$ that is directly expressed in term of the asymmetron parameters $(\mu, M, \lambda, \kappa)$,
\begin{equation}\label{radius-representation}
R_0=\frac{3\sigma}{\varepsilon} = \frac{3\sqrt2}{\mu}\times\left(\frac{\lambda}{\kappa}\right)
\end{equation}
and we take $\kappa\simeq 0.01\lambda$. In Ref.~\cite{Clements:2023bva}, Clements et.~al suggest a spherical chamber with internal radius $R_s=10\text{cm}$ (which they call $L$) and use their symmetron parameter values, $\mu = 2\times 10^{-13}\ \text{GeV}$, $M=100\text{GeV}$ and $\lambda=10^{-10}$. With this equivalent choice of parameters, we find our critical instanton bubble to have a radius,
\begin{equation}\label{critical-asymmetron-bubble-radius}
    \begin{split}
        R_0 &= \left(\frac{3\sqrt2}{2\times 10^{-13}}\right)\times (100) \ \text{GeV}^{-1} \\
        &= 2\times 10^{15}\ \text{GeV}^{-1}\\&=40\ \text{cm}. 
    \end{split}
\end{equation}
$R_0$ is of the same order of magnitude as the suggested chamber radius $R_s=10\text{ cm}$ in Ref.~\cite{Clements:2023bva}; therefore, bubble nucleation is a physically relevant phenomenon that would be observable at the scale of these vacuum chamber experiments.

$ $

To determine the value of the exponent, we use the value of $R_0$ determined in Eq. (\ref{critical-asymmetron-bubble-radius}), 
\begin{equation}
    \begin{split}
        \frac{B_0}{2\hbar} &= \frac{\pi^2 R_0^4\varepsilon}{12} \\
        &=\frac{\pi^2(\mu R_0)^4}{18\lambda}\times \left(\frac{\kappa}{\lambda}\right)\\
        &=\frac{\pi^2\times(400)^4}{(18)\times (10^{-10})}\times0.01\\
        &=4\times10^{18} 
    \end{split}
\end{equation}
thus, the critical radius of curvature, $\Lambda \sim 40\times 10^{1.74\times 10^{18}}\text{ cm}$. Thus, in the vacuum chamber in Ref.~\cite{Clements:2023bva}, we expect the nucleation rate to be dominated by bubbles forming on the edge of the vacuum chamber as opposed to the bulk of the scalar field. However, this is the leading order correction to $\Lambda$, which means we are ignoring explicit curvature corrections to the size of the vacuum chamber. In principle, neither the edge planar nor the bulk vacuum bubble would form on a time scale comparable to the age of the Universe since their Euclidean actions are of the same order of magnitude. Setting $\kappa=\widetilde{\kappa}\lambda$, we can express $\Lambda= \frac{3\sqrt{2}}{\mu\widetilde{\kappa}}\exp\left(\frac{18\pi^2}{\widetilde \kappa^3\lambda}\right)$. This means that $\Lambda$ decreases as $\lambda$ is made larger in magnitude. Similarly, as we make $\mu$ larger, $\Lambda$ decreases. 

$ $

For a chamber filled with a subcritical gas with density $\rho$, the critical radius of curvature receives a density-dependent correction as shown in Eq.~(\ref{upper-limit-screened}). When $\rho_*-\rho\ll\rho_*$, we see that the critical radius of curvature at leading order is of the same order as the critical bubble radius, $\Lambda\gtrsim R_0$. Ref.~\cite{Clements:2023bva} suggests that to observe the symmetry-breaking transition that leads to the formation of domain walls, the vacuum chamber must be filled with a gas to restore the $\mathbb{Z}_2$ symmetry of the potential. Then the gas should be released from the chamber. As the density in the chamber falls below $\rho_*$, the scalar field forms domains of the false and true vacuum. Patches within the chamber that settle in $\phi_-$ will immediately begin to proliferate bubbles in the bulk phase; however, as the density is decreased further, the bubbles preferentially form on the chamber walls.
\subsection{Applications to Cosmic Voids}
The critical radius of curvature, $\Lambda$, of a hollow spherical void can be calculated for the symmetron at the cosmological scale. As the symmetron is proposed to be a candidate for dark matter and dynamical 
dark energy in Ref.~\cite{Hinterbichler:2011ca}, such a calculation is worth considering for the asymmetron. An example of cosmological symmetron parameters $(\mu, M, \lambda)$ are presented in Ref.~\cite{farbod}. We shall again take $\kappa=0.01\lambda$ as a representative example. A cosmic void is a large-scale structure analogous to a vacuum chamber. They are defined to be a region of space in which the density of matter is $\sim10\%$ that of the average density of the Universe. We ignore the self-gravitation of the domain walls and the expansion of the Universe for the sake of simplicity.

$ $

Using the equations in Ref.~\cite{farbod} and in Appendix \ref{app:cosmological-asymmetron}, the exemplar cosmological symmetron parameters are given by, 
\begin{equation}
    \begin{split}
        \mu &\simeq 0.5\ (\text{Mpc})^{-1}\\
        M &\simeq 5.72\times 10^{15} \ \text{GeV}\\
        \lambda&\simeq 6.312\times 10^{-104}.
    \end{split}
\end{equation}
Using Eq.~(\ref{radius-representation}), we can calculate the radius of the critical bubble,
\begin{equation}
    R_0\sim 85\ \text{Mpc}.
\end{equation}
We determine that $\mu R_0 \simeq  42.5$, which enters into the Euclidean action, 
\begin{equation}
\begin{split}
    \frac{B_\text{bulk}}{2\hbar} &= \frac{\pi^2}{18\lambda}(\mu R_0)^4\times \frac{\kappa}{\lambda}\\ 
    &= \frac{\pi^2}{18\times 6.312\times 10^{-104}}\times (42.5)^4\times 0.01\\
    &\simeq 2.834 \times 10^{107}
\end{split}
\end{equation}
and so we find that at the cosmological scale, a cosmic void with radius greater than the critical radius of curvature, $\Lambda$, given by, 
\begin{equation}
    \begin{split}
        \Lambda&= 85 e^{2.834 \times 10^{107}} \ \text{Mpc}\\
        &\simeq 85\times 10^{1.23\times 10^{107}}\ \text{Mpc}
    \end{split}
\end{equation}
is the upper limit to edge bubble nucleation. Thus, it is safe to say that at the cosmological scale for a spherically symmetric cosmological void, the edge instanton is always preferred. This is significant for the assumptions of cosmological simulations like those of Ref.~\cite{farbod} where the authors randomly seed asymmetron domains in a region and perform $N$-body simulations. Our calculation shows that at least for instantons, the cosmological asymmetron shows a preference for forming true vacuum bubbles at the edge of a cosmological void. 

$ $ 

We saw previously that the semiclassical decay rate and thus the probability of bubble nucleation are maximised for a peak matter density that is very close to the $\rho_*$. The cosmological symmetron parameters in Appendix \ref{app:cosmological-asymmetron} allow us to determine the critical density of the symmetron potential to be, 
\begin{equation}
    \rho_*\sim 3.4\times 10^{-8} \ \text{g/cm}^3
\end{equation}
To determine the density of a galactic void, we would like to calculate the average matter density of the Universe. The critical density, $\rho_\text{crit}$, is given by, 
\begin{equation}
    \rho_\text{crit}=\frac{3H_0^2}{8\pi G}\simeq8\times10^{-30} \text{ g/cm}^3
\end{equation}
and the average matter density, $\rho_m$ is given by, 
\begin{equation}
    \begin{split}
        \rho_m =\Omega_m \rho_\text{crit}= 0.315\times \rho_\text{crit} \simeq 2.7\times 10^{-30}\text{ g/cm}^3.
    \end{split}
\end{equation}
Finally, a cosmic void has a matter density, $\rho_\text{void}\sim0.1 \rho_m$, which gives, 
\begin{equation}
    \begin{split}
        \rho_\text{void}\simeq 2.7\times 10^{-31}\text{ g/cm}^3.
    \end{split}
\end{equation}
which is again quite disconcerting for the bulk bubble nucleation within a cosmic void with density corrections since $1-\rho_m/\rho_*\simeq 1$. Within a cosmic void, the density corrections cannot sufficiently reduce $\Lambda$ to allow bulk nucleation. However, in more dense environments that are still subcritical, we expect the decay rate to be much larger.
\section{Concluding Remarks}
In this work, we have shown that the coupling of the asymmetron to environmental density distributions allows bubbles to nucleate in more elaborate ways than the bulk nucleation mechanism of Ref.~\cite{PhysRevD.15.2929}. We first demonstrated that dense objects immersed in a false vacuum background acquire a surface energy density $\sigma_\pm$ due to the screening mechanism of the asymmetron. For sufficiently large and dense objects, the planar limit yields a surface energy density that is determined solely by the properties of the vacuum potential, such as the external VEV and mass of the asymmetron. Using the Nambu-Goto action, we calculated the decay rate of the false vacuum in the presence of a homogeneous gas of subcritical density. The enhanced nucleation rate in such environments arises because the barrier in the effective potential is both lower and narrower in regions of higher density. As the system undergoes density-driven symmetry breaking, there is a rapid proliferation of bubbles, which diminishes as the gas continues to diffuse and the density decreases. The Nambu-Goto action was then extended to include nucleation on substrates of supercritical density. Such substrates carry a surface tension and introduce boundary terms into the effective Nambu-Goto action. For the planar substrate, we found that bubbles nucleate as spherical caps, and we paid special attention to the hemispherical bubbles, which contribute half the Euclidean action of a bulk bubble. Thus, boundary nucleation is preferred semiclassically. For cylindrical substrates, our results reproduce the earlier semiclassical analyses in Ref \cite{PhysRevD.110.105015} for convex geometries, and we extended these calculations to concave cylinders. We find that concave cylindrical substrates give the smallest Euclidean action, implying the strongest enhancement of nucleation rates. This is especially relevant for laboratory searches for fifth forces of Ref.~\cite{Clements:2023bva}, where the walls of a vacuum chamber may be modelled as concave cylindrical substrates and possible bubble formation in cosmic voids. Our results suggest that bubble nucleation is amplified in regions of high curvature, with additional enhancement from the presence of a surrounding subcritical gas.

$ $

While our analysis provides a consistent framework for nucleation catalysed by density distributions and substrates, we have limited ourselves to the hierarchy of scales $L_{\pm}\ll R_0$ and, in places, also to $R_s \ll R_0$. Firstly, our perturbative inclusion of curvature effects in Section~\ref{sec:perturbative} breaks down when the bubble radius and curvature radius of the substrate become comparable. This could be overcome by solving the Nambu-Goto equations for the worldsheet instantons numerically, perhaps extending methods developed for worldline instantons~\cite{Gould:2017fve}. Secondly, the Nambu-Goto formalism itself breaks down when the bubble wall thickness becomes comparable to its radius. While the effects of the bubble wall width can be carried out perturbatively~\cite{Megevand:2023nin, Megevand_2024, Megevand_2025}, eventually they must be included to all orders by solving the instanton field equations directly. Another limitation is the assumption of zero temperature. At low temperatures, the spherical bounces we have considered simply become periodic in imaginary time with frequency proportional to the inverse temperature, $\frac{1}{T}$, but in the high-temperature regime ($T\gg \mu$), the system undergoes dimensional reduction, with the $S^3$ bounce replaces by a cylindrical configuration $\mathbb R\times S^2$, \cite{Garriga:1994ut, PhysRevD.110.105015}. A systematic analysis of thermal nucleation rates in the asymmetron model remains an important extension. 

$ $

Finally, because the asymmetron is a scalar–tensor theory, gravitational effects are essential for a complete description. In particular, we expect that regions near the critical density (where nucleation is most prolific) could lead to collisions and coalescence of bubbles, producing gravitational wave signatures. Such processes may contribute to the stochastic gravitational wave background, offering a potential observational window into the dynamics we have described.

\section*{Acknowledgments}
CB would like to thank \textit{Alex Jenkins} for interesting conversations regarding bubble nucleation within analogue systems and pointing out the work of Canaletti and Moss in Ref.~\cite{PhysRevD.110.105015}. USA and CB would like to thank \textit{Ian Moss} and \textit{Peter Millington} for useful comments. USA would like to thank \textit{Lauren Gaughan} for assistance in producing FIG.~\ref{fig:junction-conditions-bubbles} and \textit{Joonas Hirvonen} for conversations on perturbative expansions of functional determinants. CB and PMS would like to thank the STFC for partial support under grant number ST/X000672/1. OG was supported by a Royal Society Dorothy Hodgkin Fellowship. USA was supported by the University of Nottingham. 

\section*{Data Availability}
The code developed and used to generate the numerical solutions and graphs provided can be found in the GitHub repository at: \url{https://github.com/usama-bit137/asymmetron-bubbles}.
\appendix

$ $

\section{\label{app:cosmological-asymmetron} Deriving the cosmological asymmetron parameters}
The conversion between the standard parameters of the symmetron $(\mu, \lambda, M, \kappa)$ and the cosmological parameters $(\xi_*,a_*,\beta_+, \beta_-)$ are given by the formulae presented in Ref.~\cite{farbod} as, 
\begin{eqnarray}
        L_C &=& \frac{1}{\sqrt{2}\mu} = \xi_\star\times 2998 \text{Mpc/h} \label{cosmo-compton-wavelength}\\
        M &=& M_\text{p}\xi_*\sqrt{6\Omega_{m,0}/a_*^3}\label{cosmo-mass-cutoff}\\
        \lambda &=& \frac{H_0^2}{72\Omega_m^2M_\text{p}^2}\left(\frac{a_*^3}{\beta_*\xi_*^3}\right)^2\label{cosmo-quartic-paramter}
\end{eqnarray}
where $\beta_\pm$ is the strength of the asymmetron fifth force relative to the Newtonian gravitational force on either of the scalar domains $\phi_\pm$, $\xi_\star$ is the Compton wavelength in units of the Hubble length, and $a_\star$ is the scale factor at symmetry breaking. Ref.~\cite{farbod} neglects the factors of $M_\text{p}$ in Eqs.~(\ref{cosmo-mass-cutoff}-\ref{cosmo-quartic-paramter}). Since our instanton calculations are within the thin-wall limit, it is more appropriate for us to take the symmetron convention where $\beta_\star \equiv \beta_+ = \beta_-$. A representative choice of the symmetron parameters from Ref.~\cite{farbod} is given by $(\xi_\star,  a_\star, \beta_\star) = (3.3\times 10^{-4}, 0.33, 1)$ which we can use to calculate the critical radius of curvature of a large spherical and hollow cosmological structure such as an intergalactic void. 

$ $

The symmetron mass parameter can be calculated by rearranging Eq.~(\ref{cosmo-compton-wavelength}),
\begin{equation}
\begin{split}
    \mu &= \frac{h}{\sqrt2\times \xi_*\times 2998}\ (\text{Mpc})^{-1}\\
    &\simeq \frac{0.73}{\sqrt2\times 3.3\times 10^{-4}\times 2998}\ (\text{Mpc})^{-1}\\ &\simeq 0.5 \ (\text{Mpc})^{-1}
\end{split}
\end{equation}
which is good because this results in a cosmologically large value for $L_C$. Similarly, the dimensionless quartic coupling of the symmetron is given by, 
\begin{equation}
\begin{split}
 \lambda &= \frac{H_0^2}{72\Omega_m^2M_\text{p}^2}\left(\frac{a_\star^3}{\beta_\star\xi_\star^3}\right)^2\\
 &= \frac{(8.761\times 10^{-61})^2(0.73)^2}{72(0.3)^2}\left(\frac{(0.33)^3}{(3.3\times10^{-4})^3}\right)^2\\
 &\simeq 6.312\times 10^{-104} 
 \end{split}
\end{equation}
where we have used the value of $H_0 = 2.1332 h\times10^{-42}\ \text{GeV}$ from Ref.~\cite{Kolb:1990vq}. The symmetron mass cut-off, $M$, is given by, 
\begin{equation}
\begin{split}
    M &\simeq 3.3\times 10^{-4}\times M_\text{p}\times \sqrt{6\times 0.3/(0.33)^3}\\
    &\simeq 5.72\times 10^{15} \ \text{GeV}.
\end{split}
\end{equation}
\bibliography{apssamp}

\end{document}